\documentclass{aa}
\usepackage{graphicx}
\newcommand{\fma}[1]{\mbox{$#1$}}

\newcommand{\unit}[1]{\ifmmode \:\mbox{\rm #1}\else \mbox{#1}\fi}
\newcommand{\mone}{\fma{^{-1}}}

\newcommand{\etal}{{et al.\/}}
\newcommand{\eg}{{e.g.\/}}

\newcommand{\ie}{{i.e.\/}}

\newcommand{\ha}{H$\alpha$}

\newcommand{\hd}{H$\delta$}
\newcommand{\hi}{H~{\sc i}}
\newcommand{\hii}{H~{\sc ii}}
\newcommand{\hei}{He~{\sc i}}
\newcommand{\heii}{He\,{\sc ii}}

\newcommand{\Ni}{[N~{\sc i}]}

\newcommand{\nii}{[N~{\sc ii}]}

\newcommand{\sii}{[S~{\sc ii}]}

\newcommand{\oii}{[O~{\sc ii}]}
\newcommand{\oiii}{[O~{\sc iii}]}
\newcommand{\siii}{[S~{\sc iii}]}
\newcommand{\neiii}{[Ne~{\sc iii}]}
\newcommand{\nai}{Na~{\sc i}}

\newcommand{\feiii}{[Fe~{\sc iii}]}

\newcommand{\kms}{\unit{km~s\mone}}

\newcommand{\msun}{\unit{M$_\odot$}}

\begin{document}

   \title{Dynamical masses of two young globular
      clusters in the blue compact galaxy
      ESO\,338--IG04\thanks{Based on observations collected at
European Southern Observatory, Paranal, Chile, under observing programme
65.N-0704; and on observations with the NASA/ESA {\it Hubble Space
Telescope}, obtained at the Space Telescope Science Institute, which is
operated by the Association of Universities for Research in
Astronomy, Inc., under NASA contract NAS5-26555. }}

   \author{G{\"o}ran {\"O}stlin\inst{1}
          \and
          Robert J. Cumming\inst{1}
          \and
          Nils Bergvall\inst{2}
          }

   \offprints{G.\ {\"O}stlin}
 
   \institute{Stockholm Observatory, AlbaNova University Center, Stockholms
           Center for Physics, Astronomy and Biotechnology, Roslagstullsbacken
                           21, SE-106 91 Stockholm, Sweden.\\
              \email{ostlin@astro.su.se, robert@astro.su.se}
          \and
	  Uppsala Astronomical Observatory, Box 515, SE-751 20
	Uppsala, Sweden.}
   \date{Received ; accepted 26 September 2006}

   \abstract{ 

We present high-resolution {\'e}chelle spectroscopy, obtained with the
UVES spectrograph on ESO/VLT, of two luminous star clusters in the
metal-poor blue compact galaxy ESO\,338--IG04 at a distance of 37.5
Mpc. Cross-correlating with template stars, we obtain line-of-sight
velocity dispersions of 33 and 17 \kms. By combining with size
estimates from Hubble Space Telescope images we infer dynamical masses
of $1.3\times10^7$ \msun\ and $4.0\times10^6$ \msun for the two
clusters, making them among the most massive known. 
The less massive cluster is 
the faintest cluster for which a dynamical mass has yet been
obtained. In both clusters we detect Balmer absorption lines which we
use to estimate their ages.  From the younger ($\sim 6$ Myr) and more
massive cluster, we detect \heii\ $\lambda$4686 emission of
intermediate width, indicating the presence of very massive O-stars.
Moreover, analysis of the \oiii\ $\lambda$5007 and \ha\ emission lines
from the region near the younger cluster indicates that it is
associated with a bubble expanding at $\sim40$ \kms. We also see from
the \nai\,D absorption lines indications of neutral gas flows towards
the younger cluster.  We compare the dynamical masses with those
derived from photometry and discuss implications for the stellar
initial mass function (IMF) in each cluster. Both clusters are
compatible with rather normal IMFs which will favour their long-term
survival and evolution into massive bona fide globular clusters.

   \keywords{galaxies: evolution -- galaxies: individual: ESO338--IG04
                (= Tol 1924-416) --
                galaxies: starburst -- galaxies: star clusters --
                galaxies: stellar content
               }
   }
   \titlerunning{Dynamical masses of globular clusters in ESO338--IG04}
   \authorrunning{{\"O}stlin et al.}
   \maketitle

   \begin{figure*}
% \resizebox{\hsize}{!}{\includegraphics{uves_acquisition2.eps}}
 \resizebox{\hsize}{!}{\includegraphics{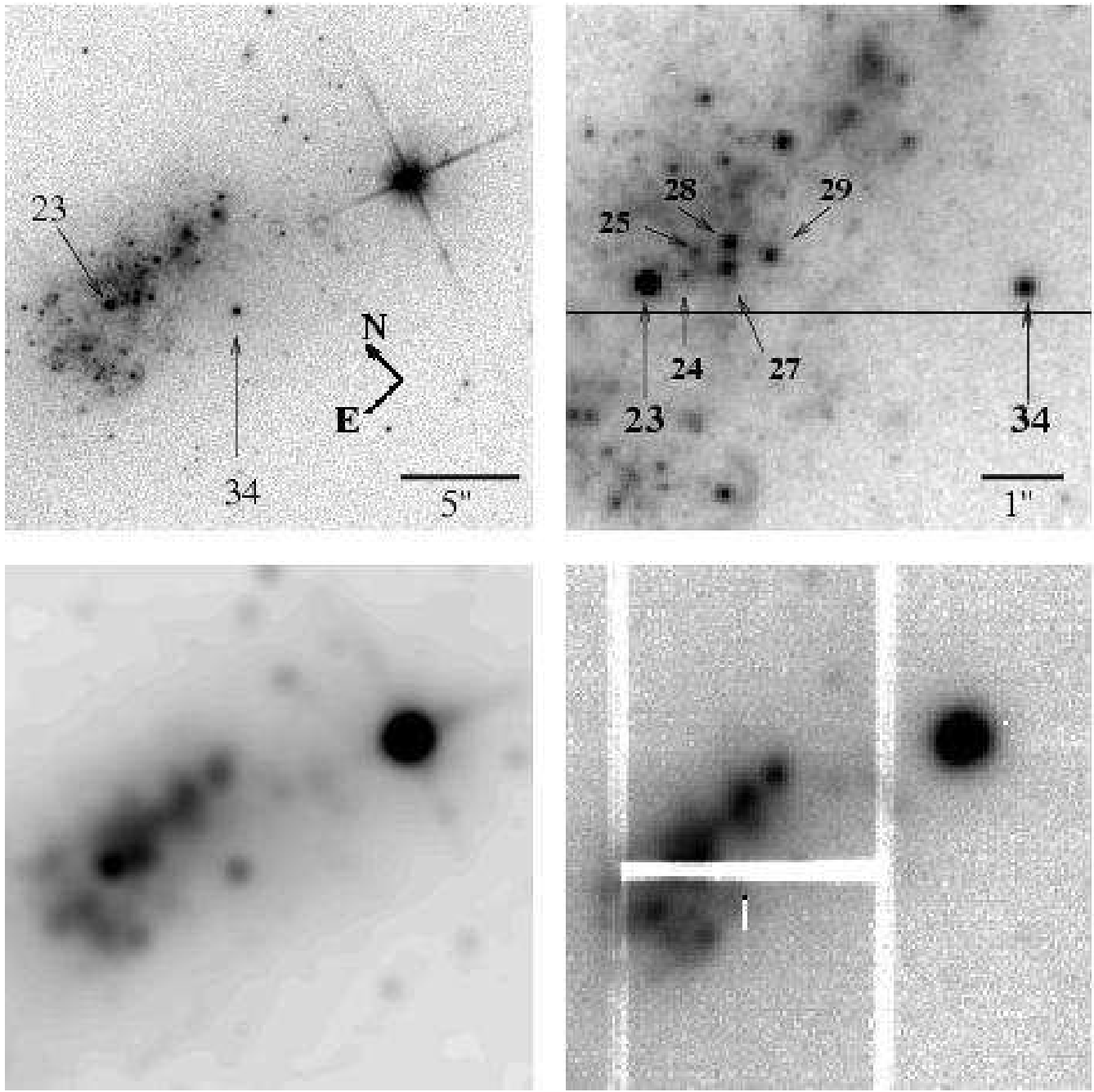}}
      \caption{Acquisition. {\bf Upper left}: HST/WFPC2  image of ESO\,338--IG04 in the F814W band, 
                rotated to match the position angle used for our UVES
                observations. The orientation is indicated and the two
                main targets for the spectroscopy discussed in this
                paper are labelled.
		{\bf Upper right}: Enlargement
		of the upper left image showing the slit location, the
		main targets and some other nearby clusters.  The
		width of the slit is $0.7\arcsec$.  The clusters have
		been identified with the index number used by {\"O}stlin
		\etal\ (\cite{OBR}).  All but \#34 belong to their
		`inner sample'.
		{\bf Lower left}: The image in the upper left panel,
		convolved with a Gaussian kernel the same size as
		the seeing (FWHM=0\arcsec.65) to simulate the actual
		observing conditions.
		{\bf Lower right}: Slit view image from the red camera
		of UVES. The slit is the horizontal bright bar, and
		has a length of 11\arcsec.  The bright foreground
		star on the right was used for offsetting in the
		acquisition.} 
\label{acq} 
\end{figure*}

\section{Introduction}

Super star clusters (SSCs) are important sites for star formation in
starburst galaxies (Arp and Sandage \cite{arp}, Meurer \etal\
\cite{meurer}), and may be the progenitors of globular clusters (GCs).
Different observational techniques have been used to investigate the
nature and history of these objects. High resolution imaging, notably
with the Hubble Space Telescope (HST), has been used to collect
photometry and to estimate the sizes of many SSCs (Whitmore
\cite{whitmore_rev}).  Photometry, in combination with spectral
evolutionary synthesis models, has also been used to estimate cluster
ages and masses ({\"O}stlin \etal\ \cite{OBR}).  While many such {\em
photometric} masses have been determined, only a small fraction of
extragalactic SSCs and GCs have {\em dynamical} mass estimates.  Taken
together, measurements of the photometric and dynamical masses can allow
constraints on the stellar initial mass function (IMF) and the
dynamical state in these objects.

The spectrum of a cluster can provide other information about the
stellar population.  For example, Gonz{\'a}lez Delgado \etal\
(\cite{GD}) inferred a population of Wolf-Rayet stars in the massive
cluster NGC 1569-A, and Origlia et al.\ (\cite{O01}) used ultraviolet
spectra to show that the same cluster has a population of O-stars.

High-resolution spectra can be used to measure the velocity
dispersion, and, assuming virialisation, to estimate dynamical masses
for SSCs and GCs.  Ho and Filippenko (\cite{HFa}, \cite{HFb}) were
first to use the cross-correlation method described by Simkin
(\cite{Simkin}) and Tonry and Davis (\cite{TD}) to measure the velocity
dispersions of NGC 1569-A (giving a kinematic mass of $3.3\times10^5$
\msun) and the SSC in NGC 1705 ($8.2\times10^4$ \msun).  The same
method has been applied to M82-F by Smith and Gallagher (\cite{SG};
$1.2\times10^6$ \msun), to five clusters in NGC 4214 and NGC 6946 by
Larsen \etal\ (\cite{Larsen01}, \cite{Larsen04}; $0.2-1.8\times10^6$
\msun), by Gilbert and Graham (\cite{GG}) to clusters A1, A2 and B in
NGC 1569 ($3.9\times10^5$ \msun, $4.4\times10^5$ \msun \ and
$2.3\times10^5$ \msun, respectively) and by McCrady, Gilbert and Graham
(\cite{MGG}) to two SSCs in M82 ($1.5\times10^6$ \msun\ and
$3.5\times10^5$ \msun).  Mengel et al.\ (\cite{M}) measured the
dynamical masses of five SSCs in NGC 4038/4039 ($0.65 - 4.7\times10^6$
\msun) by fitting model spectra in the near-IR and optical.  Maraston
et al. (2004) presented a dynamical mass estimate of as much as
$8\times10^7$\msun\ for the cluster W3 in NGC~7252, which may
possibly be a dwarf galaxy.  Bastian \etal\ (\cite{Bastian}) found
masses of $1.6\times10^7$ \msun\ for both W30 in NGC 7252 and G114 in
NGC 1316, and present a compilation of previously published 
results (their Table 6).

A possibly related class of objects are the nuclear star clusters
found in many bulgeless spiral galaxies (e.g. Walcher \etal\
\cite{walcher}), some of which appear to be very massive but which
probably have more extended star formation histories than GCs and
SSCs.

ESO\,338--IG04, also known as Tol\,1924-416, is a luminous ($M_B=-19$)
blue compact galaxy (BCG) --- the closest in a class of
galaxies that are rare in the local universe but that become
increasingly important towards higher redshifts and the peak in the
cosmic star formation rate at $z\sim 1$ (\eg\ Werk \etal\ \cite{Werk}).
ESO\,338--IG04 shows both vigorous star formation and a rich
population of more than 100 young SSCs and old globular clusters
({\"O}stlin \etal\ \cite{OBR})\footnote{Following {\"O}stlin \etal\
(\cite{OBR}) we adopt a distance of 37.5 Mpc to ESO\,338--IG04.}.
Analysis of the age distribution of these clusters shows evidence that
at least one strong starburst occurred in the past, a couple of Gyr
ago, and the present starburst has probably been active for about 40
Myr ({\"O}stlin \etal\ \cite{OZBR}). In addition ESO\,338--IG04 hosts a
population of cosmologically old (age $\le 10$~Gyr) globular
clusters. What makes ESO\,338--IG04 particularly interesting as a star
cluster formation laboratory is its combination of small reddening,
many clusters, and a low metallicity ($12+\log({\rm O/H})=8.0$).
Compared to other metal-poor star-forming galaxies in the local
universe (such as NGC\,1569, NGC\,1705, and He\,2-10) the clusters in
ESO\,338--IG04 are more luminous and numerous.  The present starburst
has been triggered by a small merger or interaction with its companion
galaxy ({\"O}stlin \etal\ \cite{ostlin01}, Cannon \etal\ 2004). Hence
ESO\,338--IG04 offers the possibility of directly studying globular
cluster formation associated with hierarchical galaxy evolution, at a
low metallicity.

In this paper, we report high-dispersion {\'e}chelle spectroscopy of two
of these clusters, labelled Inner-\#23 and Outer-\#34 by {\"O}stlin et al.
(\cite{OBR}).  Hereafter, we refer to them simply as \#23 and \#34.
Cluster \#23, a young blue cluster situated in the centre of the
starburst region, is the most luminous in the whole galaxy
($M_V=-15.5$).  Cluster \#34 is of intermediate age, $\sim 1$ Gyr, but
still very luminous ($M_V=-12.8$), which led {\"O}stlin \etal\
(\cite{OBR}) to suspect a mass in excess of $10^7$ \msun.  Here, we
use high-dispersion VLT/UVES spectroscopy to make dynamical mass
estimates and new age estimates for both clusters. These two
clusters were selected since they represent the most luminous of the
young and intermediate age clusters, respectively, and could be
observed in a single UVES slit.

The outline of the rest of the paper is as follows. In Sect.\ 2
we describe the observations and reductions.  In Sect.\ 3 we show the
results obtained; subsection 3.4 in particular describes how we
obtained estimates of the velocity dispersions of the clusters through
cross-correlation techniques. In Sect.\ 4 we discuss the inferred
dynamical masses and compare with results derived from photometry.
Sect.\ 5 summarises our conclusions.

\section{Observations and reductions}

\subsection{UVES-spectroscopy}

Our spectra were taken on 2000 June 15 using the {\'e}chelle spectrograph
UVES mounted on the telescope Kueyen (UT2) of the Very Large
Telescope, at the European Southern Observatory's Paranal site. The
observations were performed in service mode.  The seeing was typically
0$\arcsec$.6.

We placed the slit at a fixed position angle of 42$^{\circ}$.6 in
order to capture the spectra of both cluster \#23 and \#34 in the same
exposure (see Fig. \ref{acq}).  To minimise contamination from other
sources in this crowded starburst environment, we used a slit width of
0\arcsec.6.  
The slit lengths used were 8\arcsec\ and 11\arcsec\ for the blue and red
cameras, respectively.  The atmospheric dispersion corrector in UVES
was used to minimise slit losses due to differential refraction.

Using the dichroic setting DIC1 with cross-dispersers CD2 and CD3, we
covered the optical spectrum from 3280\AA\ to 6660\AA\ at a resolution
of $R\sim 60000$.
We binned the CCD pixels $2\times 2$ to lower the influence of the
readout noise, which resulted in a spatial scale of 0\arcsec.50 for
the blue arm and 0\arcsec.36 per pixel for the red arm.  A total of
eight 2000-second exposures were taken.  The data were reduced using
the UVES pipeline, with extraction and sky windows modified to allow
for the extraction of two non-centred sources.  Our wavelength
calibration, done by comparison with ThAr lamps, should be accurate to
better than 0.02 \AA, or about 1 \kms.

Using the same instrumental set-up, we observed five template stars
(for details, see Table \ref{tab-templates}).  Two of these, HR 7706
and HR 7749, were observed on 2000 April 22; the rest were observed on
the same night as the cluster observations.

\begin{table*}
\begin{center}
\caption[]{Template stars}
\label{tab-templates}
\begin{tabular}{llcl}
\hline
Designation & Type and class & Radial velocity (\kms) & Reference$^{\rm 1}$ \\
\hline
HR~6961 & K4\,{\sc i}{\small b} & $-2.291$ & Nidever \etal\ (\cite{Nidever})\\
HR~7277 & K1\,{\sc i}           & $+2.8$ & Ochsenbein (\cite{Ochsenbein}) \\
HR~7706 & K1\,{\sc iii}         & $+1.0\pm0.3$ & This work \\
HR~7749 & F5\,{\sc v}           & $-30.8$ & Kharchenko \etal\ (\cite{Kharchenko})\\
HR~8299 & G5\,{\sc iii}         & $-57.1\pm0.4$ & This work \\
\hline

      \end{tabular}
\end{center}
\noindent Note: 
(1) All references are to catalogues available at the VizieR website, http://vizier.u-strasbg.fr/ (Ochsenbein, Bauer and Marcout \cite{VizieR}).
    \end{table*}

\subsection{Source extraction}\label{sec-spatial}

Special care was taken with the background subtraction, though the
small size of the slit and the complex \hii\ region emission meant
that the subtraction is unreliable in the vicinity of strong emission
lines.  This is also due to the intrinsically different spatial
distribution of nebular emission and the stellar continuum (see {\"O}stlin
\etal\ \cite{OZBR} and Fig.\ \ref{o3}).  Spectra were extracted using
average extraction, centred on the continuum of the clusters.  The
extraction windows were 3 and 5
pixels wide for the blue and red arm spectra, 
respectively.  We extracted the spectra for both the clusters and 
the template stars in this way.

While clusters \#23 and \#34 are well-separated on the sky, light from
a number of other sources was also collected by our slit (see Fig.\
\ref{acq}).  In particular, we expect a contribution from at least two
more clusters from the inner sample of {\"O}stlin \etal\ (\cite{OBR}): numbers
24 and 27 (\#24 and \#27, see Fig.\ \ref{acq}).  Both are
slightly offset from the centre of the slit and
we estimate that about 50\% of the flux from \#24 and about 15\% of
that from \#27 are collected by our extraction window.  Both clusters
are however fainter than \#23, by 4.2 and 2.1 magnitudes,
respectively, in the F555W filter ({\"O}stlin et al.\ \cite{OBR}).  Based
on this
we estimate that on the order of 1\% or less of the flux from
\#23 has its origin in each of \#24 and \#27. Contamination from
known sources should thus not be a problem.

Nevertheless, since \#23 is located in a crowded area, we also tried
using narrower extraction windows (width 1 pixel for the blue arm,
and 3 pixels for the red arm) with the centre offset by 1 pixel to the
north-east, where the crowding is less severe (see Fig.\ \ref{acq}).

\subsection{HST Imaging}\label{imaging}

In addition to the UVES spectra, we used images from the Hubble Space
Telescope  obtained with the Planetary Camera (PC) aperture of the
WFPC2 in the F218W, F336W, F439W, F555W and F814W passbands. These
data are described fully in {\"O}stlin \etal\ (\cite{OBR} and \cite{OZBR}).  We used
these images to estimate the sizes of the two clusters, and to
constrain their age and photometric mass from comparison with spectral
evolutionary synthesis models. The photometry used in the current
paper differs slightly from that of {\"O}stlin \etal\ (\cite{OBR}) in that 
we have optimised the photometric parameters specifically for these
two clusters and also corrected for the charge transfer (in)efficiency
of WFPC2 (Whitmore \etal\ \cite{Whitmore}, Dolphin \cite{dolphin}) using the web-based 
tool CTE Tool \#1\footnote{CTE Tool \#1 by Andrew Dolphin \\
www.stsci.edu/instruments/wfpc2/Wfpc2\_cte/wfpc2\_cte\_calc.html}.
This effect is noticeable for faint sources on low backgrounds and is
here important mainly for the F336W magnitude of \#34 where 
it amounts to $\sim$0.06 magnitudes.

\section{Results}

\subsection{The spectrum of cluster \#23}
\label{spec23}
   \begin{figure} \resizebox{\hsize}{!}{\includegraphics[bb=1 150 580
   430]{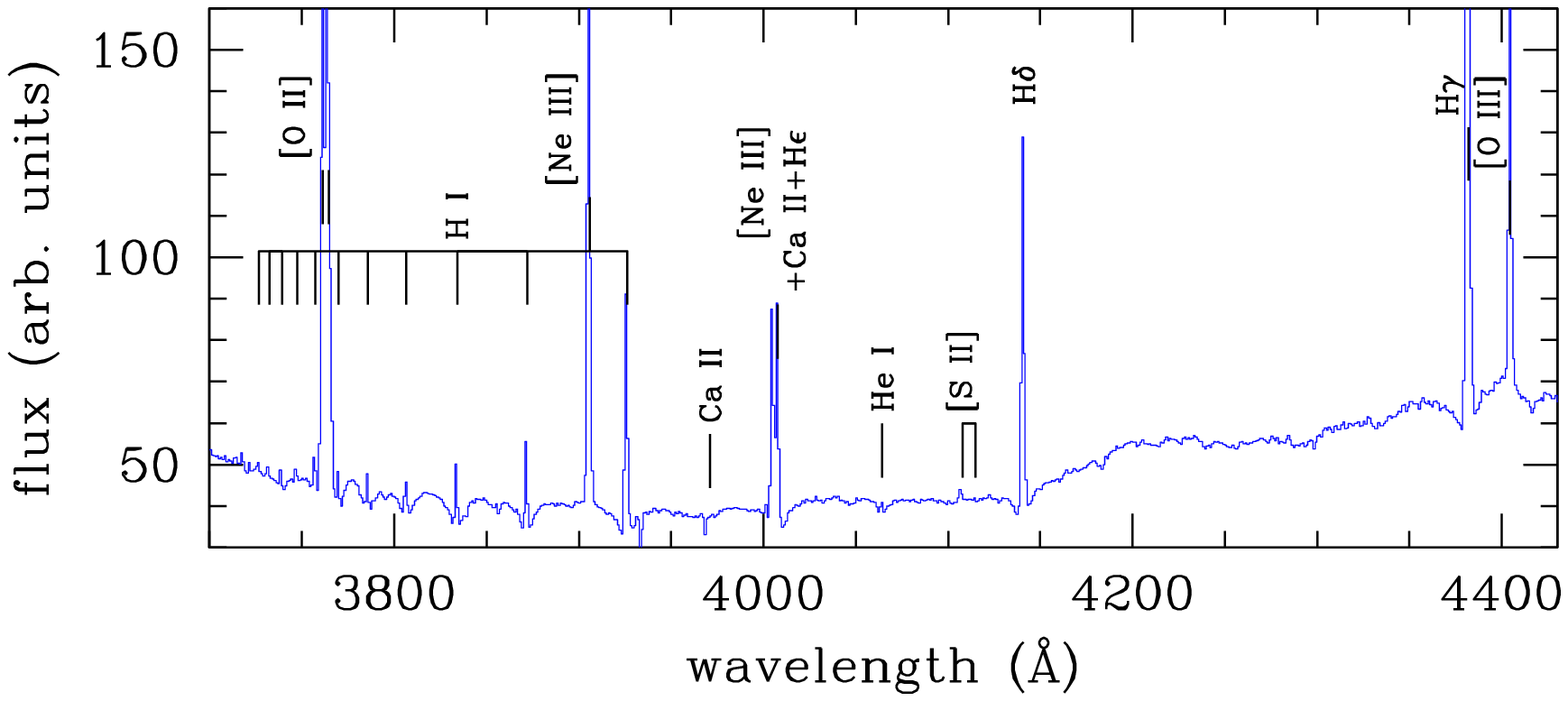}} \resizebox{\hsize}{!}{\includegraphics[bb=1 150
   580 400]{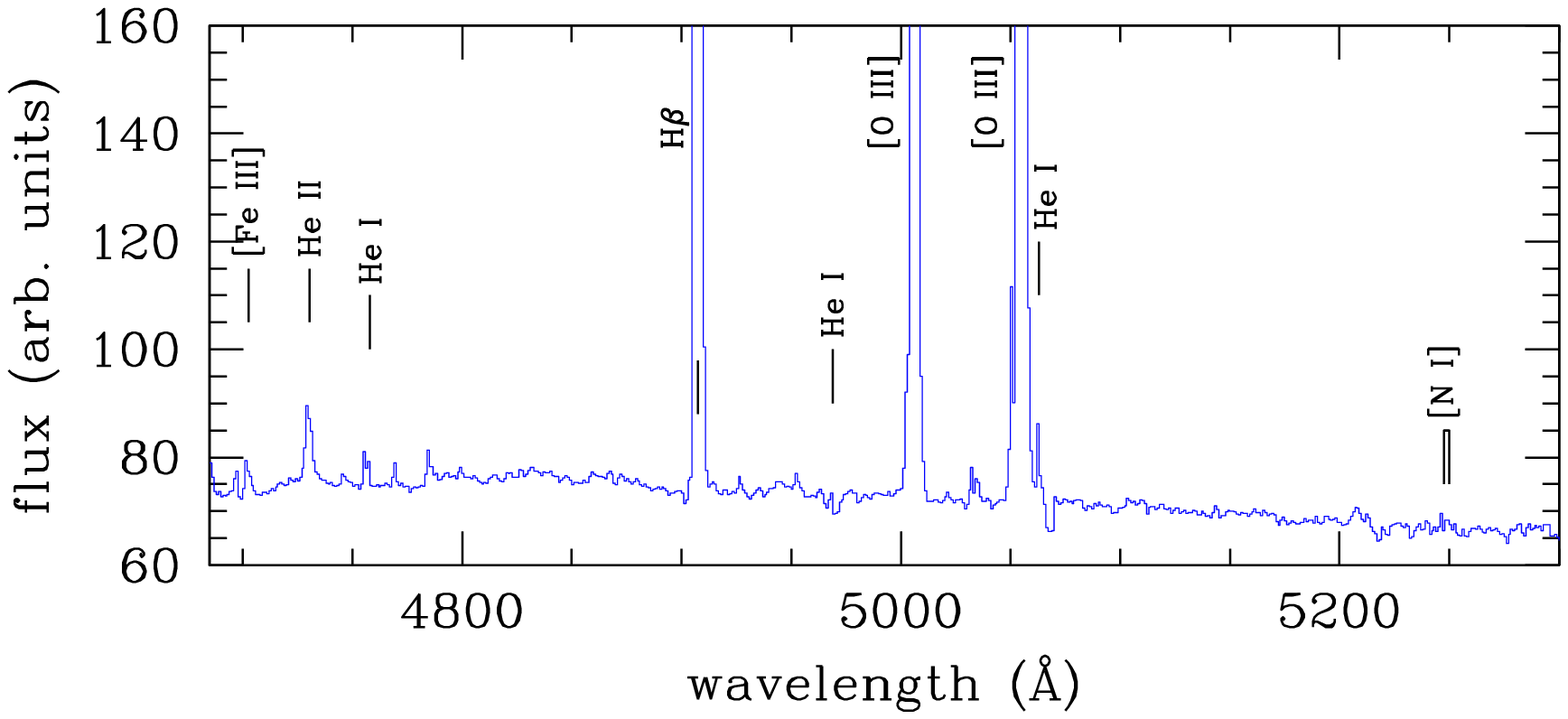}}
      \caption{Part of the spectrum of cluster \#23 with line
               identifications.  The spectrum has not been flux-calibrated.  
               Note the absorptions in the Balmer series
               and the broad emission line in \heii\ $\lambda4686$.} 
   \label{f-spec-23B} 
   \end{figure}

Figure \ref{f-spec-23B} 
shows parts of the spectrum of cluster \#23.  The spectrum shows a
stellar continuum with superimposed narrow emission lines and a few
emission lines of intermediate width.  The narrow emission lines come
from extended emission close to the cluster.  Background subtraction
across this emission was not always successful, leading to the
spurious absorption features seen in some of the emission lines.

The narrow line emission presumably originates from a combination of
\hii\ regions and supernova remnants.  We have identified lines of
\hi, \hei, \Ni, \nii, \oii, \oiii, \neiii, \sii, \siii\ and \feiii.
The ratio of \oii\ $\lambda3729$ to $\lambda3726$ indicates a emitting
electron density of around 150 cm$^{-3}$.  

In addition, broader emission is seen in \heii\ $\lambda$4686 (${\rm
FWHM}\sim460$ \kms) and possibly C~{\sc iv} $\lambda$4658 (blended
with narrower \feiii).  The line widths we see here are comparable
to those seen in Of stars, but narrower than in Wolf-Rayet stars (cf.\
Nota \etal\ \cite{N96}).

Stellar absorption lines with broad wings are seen in the Balmer
series.  Narrower absorptions are seen in \hei, \nai~D (both at the
redshift of ESO\,338--IG04 and at zero redshift from the local Galactic
ISM) and Ca~{\sc ii}~K.  A number of other absorption features are
apparent towards the red parts of the spectrum, in particular at
5168/5172 \AA\ (Mg~{\sc i}), 5185 \AA\ (Mg~{\sc i}, MgH), and also at
6138/42 \AA\ and 6496 \AA .

\subsection{The spectrum of cluster \#34}

Cluster \#34 was detected with much poorer signal-to-noise ratio than
\#23.   Its spectrum shows \nai~D absorption lines, the Mg~{\sc i} doublet at
5168/5172 \AA, and a number of other features also seen in the
template stars.  No broad emission lines of the type seen in cluster
\#23 were detected, though features with the same equivalent widths (EW)
could easily be hidden in \#34's noisier spectrum.

\subsection{Cluster ages from Balmer lines}
\label{balmer}

We have estimated the ages of the clusters by fitting the model
spectra of Gonz{\'a}lez Delgado and Leitherer (\cite{GDL99}) to the Balmer
and \hei\ line spectrum around \hd\ and in the region from H8 to H11.

First we smoothed our spectra to match the resolution of the models.
Comparing the models with the smoothed data, we located the
combination of effective temperature and surface gravity which gave
the best fit to the observed line spectrum.  Since the line cores are
dominated by emission in our data, we matched based on the wings of
the lines.
  
\begin{figure}
 \resizebox{\hsize}{!}{\includegraphics{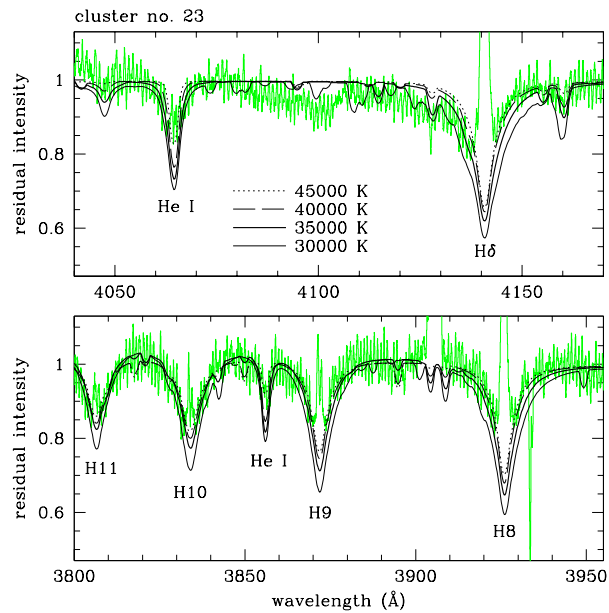}}

      \caption{Normalised spectrum of cluster
      \#23, with superimposed models of Gonz{\'a}lez Delgado and Leitherer
      (\cite{GDL99}).  We show models for $\log g=4.5$ and effective
      temperatures of 30\,000 K (thin solid line),
      35\,000 K (thick solid line), 
      40\,000 K (broken line)
      and 45\,000 K (dotted line). 
      For ease of comparison, the slope of the model spectra
      shortward of 3940\AA\ has been adjusted to match the data.
}
      \label{f-age-23} 
   \end{figure}

\begin{figure}
 \resizebox{\hsize}{!}{\includegraphics{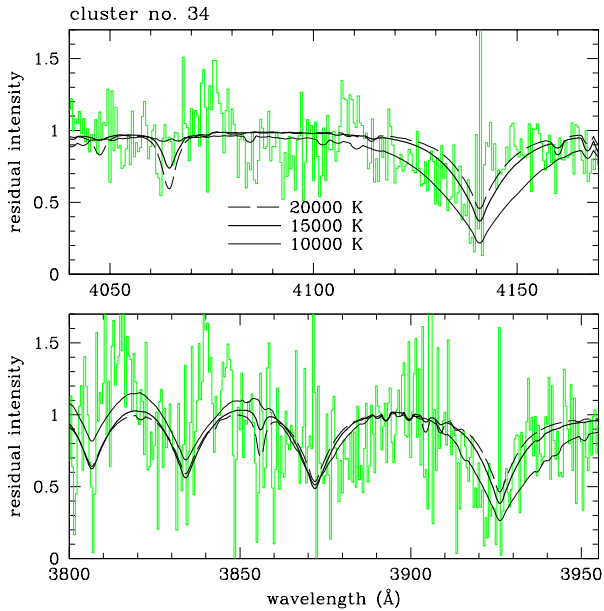}}
      \caption{As Figure \ref{f-age-23}, but for cluster \#34.
      Here the models have effective temperatures of 10\,000 K (thin solid
      line),
      15\,000 K (thick solid line)
      and 20\,000 K (broken line).
      } 
   \label{f-age-34} 
   \end{figure}

For cluster \#23 we get $T_{\rm eff}=3.75\pm0.5\times10^4$ K
(corresponding to late O stars), log~$g$ = 4.5$-$5; and for \#34 we
get $T_{\rm eff} = 1.3^{+0.7}_{-0.3}\times10^4$ K (late B stars),
log~$g$ = 4.5$-$5.  In Figures \ref{f-age-23} and \ref{f-age-34} we
show some of these fits.  Adopting log~$g$ = 4.5, we estimated the
equivalent widths of the \hd\ and H8 lines using the data in Tables 4
and 5 of Gonz{\'a}lez Delgado and Leitherer (\cite{GDL99}, GDL99).  We obtain, for
\#23, $W_{\rm H\delta} = 3.6^{+1.5}_{-0.6}$ \AA\ and $W_{\rm
H8}=2.8^{+0.7}_{-0.4}$ \AA, and for \#34, $W_{\rm H\delta} =
12^{+6}_{-4}$ \AA\ and $W_{\rm H8}=9^{+3}_{-3}$ \AA.

Finally, we estimated the age of each cluster by comparing these
equivalent widths derived from \hd\ and H8 with the values for
different instantaneous population ages presented by GDL99 (their
Table 5, for $Z$=0.001 and a Salpeter (\cite{salpeter}) IMF\footnote{We
parameterise the IMF as a power law, $dN/dM \propto M^{-\alpha}$,
where a Salpeter IMF has $\alpha=2.35$} with mass range 1 to 80
\msun).  We obtain $\log t\rm{(yr)} = 6.8\pm0.2$ for cluster \#23, or
$6^{+4}_{-2}$ Myr.

This metallicity is slightly lower than the ISM abundance of
ESO\,338--IG04, which could lead us to overestimate the age slightly.
Moreover, one may speculate that the high dynamical mass of \#23 (see
Sect.\ 3.7) could allow some self-enrichment with the result that this
cluster would be richer in metals than its environment. On the other
hand, the use of an upper mass limit of 80 \msun, as in the GDL99
model, would lead to an underestimate of the age if the true IMF
were to extend to 120 \msun.

For the fainter cluster \#34, we cannot set very stringent limits on
the age.  One reason is the noisy spectrum, leading to uncertain EW
estimates.  In addition, the Balmer line absorption EW of a
low-metallicity stellar population increases with time up to an age
$\sim 1$ Gyr, after which it decreases again (Bica and Alloin
\cite{bica}).  Based on the GDL99 models, the estimated EW indicates
an age $\sim 0.3$ Gyr, with a lower limit on the age of 50 Myr.
Comparing also with the empirical \hd\ data from Bica and Alloin
(\cite{bica}), we find for \#34 an allowed age range from about 0.5 to
3 Gyr.

As a consistency check, we compared the effective temperature of the
best-fitting models to the effective temperature at the main-sequence
turnoff in the models of Bertelli et al. (\cite{Bertelli}).  Stars at
the turnoff are expected to dominate the blue light from the clusters.
This estimate gives $\log t\rm{(yr)} = 6.8\pm0.25$ for cluster \#23.
For \#34 we obtain  $\log t\rm{(yr)} = 8.6\pm0.6$  for a low 
metallicity ($Z=0.001$) and $\sim 0.5$ dex lower for a high metallicity
($Z=0.02$). 

Thus we find an age of $\sim 6$ Myr for cluster \#23, and a rather
generous age span for \#34.
These results are consistent with our photometric modelling ({\"O}stlin
\etal\ \cite{OZBR}), which we discuss in Sect.\ \ref{photmass}.


   \begin{figure}
 \resizebox{\hsize}{!}{\includegraphics{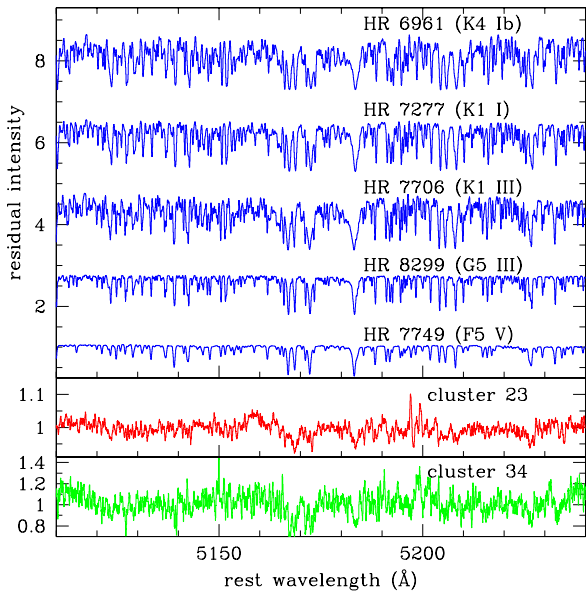}}
      \caption{Part of the spectra of the two clusters, showing the 
      region around the Mg~{\sc i} triplet, compared to the same
      region in the five template stars.  The spectra have been
      smoothed with a 9-pixel boxcar and shifted to a rest wavelength
      scale for clarity of presentation.}
   \label{f-refstars} 
   \end{figure}

\subsection{Cross-correlation analysis: 
 radial velocities and line-of-sight velocity dispersions}
\label{sigma}

To estimate the radial velocities and line-of-sight velocity
dispersions of the two clusters, we cross-correlated their spectra
with a number of template stars, following the method of Simkin
(\cite{Simkin}), Tonry and Davis (\cite{TD}) and Ho and Filippenko
(\cite{HFa}, \cite{HFb}). For this we used the {\sc
iraf}\footnote{IRAF (Image Reduction and Analysis Facility) is
distributed by the National Optical Astronomy Observatories, which are
operated by the Association of Universities for Research in Astronomy,
Inc., under cooperative agreement with the US National Science
Foundation.} task {\tt fxcor}.

\subsubsection{Selecting the proper template stars}
\label{refstars}

For the cross-correlation method to yield trustworthy measurements of
the velocity dispersion, the absorption line spectrum should be
dominated by stars with intrinsically narrow lines common to both the
target and the template star. Cool giants and supergiants are likely
to be the best templates.

At the age of cluster \#34, we expect the luminosity to be
dominated by late-type giants, which suggests that the optimum
template star should be of this type. Cluster \#23 is much younger,
but following the discussion in Ho and Filippenko (\cite{HFa}), we
still expect its absorption line spectrum in the visual wavelength
region to be dominated by red supergiants.  Examination of the high
resolution spectra of local stars presented by the UVES Paranal
Observatory Project ({\sc uvespop}; Bagnulo et al.\ \cite{Bagnulo})
confirms that in our wavelength range, only stars of spectral type
later than A are likely to contribute significantly to the observed
absorption line spectrum.

As an independent quantitative check, we used Starburst99
(Leitherer \etal\ \cite{sb99}; see below) to calculate the
contributions from stars of different spectral types to some of the
well-known Lick indices (Worthey \etal\ \cite{Worthey}).  Although
these are defined for resolutions much lower than ours, the type of
stars that dominate a particular index should also dominate the
individual lines that make up the index.  The cross-correlation
method should also be sensitive to these lines.  Equally, a spectral
type that gives a negligible contribution to the indices in the visual
wavelength region is unlikely to contaminate the cross-correlation
signal.

At the Starburst99
website\footnote{http://www.stsci.edu/science/starburst99/} it is
possible to run simulations that output the number of stars for each
spectral type and luminosity class, as a function of age.  We ran a
few such models (see Table \ref{sb99}) and used this information,
together with the luminosity and temperature relations from Drilling
and Landolt (\cite{DL}) and the fitting functions of Worthey \etal\
(\cite{Worthey}), to predict the relative contribution to the Lick
indices from stars of different spectral types as a function of
stellar population age.  Model G1 has the same tracks and metallicity
as the standard model used in {\"O}stlin \etal\ (\cite{OZBR}) for
estimating stellar populations in the clusters in ESO338--IG04.
Models P1 and P2 bracket model G1 in metallicity, and also bracket the
observed nebular abundance for ESO\,338--IG04.  Finally, model P3 has
a lower upper mass limit; in Sect.\ \ref{23} we discuss the
possibility of an IMF devoid of very massive stars. All models
use a lower mass limit of $0.1$ \msun\ and an IMF with slope
$\alpha=2.35$.

\begin{table}
\centering
\caption[]{Starburst99 models used to test which stellar types dominate 
the absorption line features in the spectral region 4600-6540 \AA.}
\label{sb99}
\begin{tabular}{llll}
\hline
Model & Tracks & Z & $M_{up}$ \\
& & & (\msun ) \\
\hline
P1 & Padova & 0.004 & 120 \\
P2 & Padova & 0.0004 & 120 \\
P3 & Padova & 0.004 &  20 \\
G1 & Geneva & 0.001 & 120 \\
\hline					   
\end{tabular}
\end{table}

We found that for all ages, only spectral types later than A
contribute to the indices in the spectral region 4600-6540 \AA.
For models P1, P2 and G1 and ages younger than 4 Myr, these
indices are dominated by main sequence stars of class F, G and K.
Later, red supergiants (RSGs) take over completely, and only after 15
Myr do giants and supergiants of spectral type earlier than K
contribute.  If the cluster has an IMF lacking very massive stars, as
in our model P3, the situation is somewhat different.  Here the main
sequence stars of types F to K dominate longer, up to ages of just
over 10 Myr, after which RSGs take over.  This is understandable since
less massive stars take longer to evolve off the main sequence.
Hence, if the high-mass IMF is truncated at $M_{up} = 20$\msun, our
F\,{\sc v} template star would be the best choice for
cross-correlation.  However, a low-resolution long-slit spectrum of
\#23 taken with VLT/FORS2 (Cumming et al.\ 2006, in prep.) shows the
presence of the Ca~{\sc ii} near-IR triplet, with an equivalent width
of several \AA.  This indicates that red giants or supergiants must be
present (D{\'\i}az \etal\ \cite{Diaz}). The Starburst99 models also predict
that strong Ca~{\sc ii} triplet absorption lines develop when RSGs
appear.

In summary, even though \#23 is young, we can still safely assume
that red supergiants dominate the absorption features we see, and that
such stars should be used as template stars for cross-correlation.
  
Cluster \#34 is somewhere in the range 0.3 to 1.4 Gyr old,
depending on metallicity (see Sect.\ \ref{balmer} and \ref{34}). For
such ages, our analysis using Starburst99 indicates that the
absorption features in the visual wavelength region are dominated by
red giants of luminosity classes {\sc ii} and {\sc iii}, with a
contribution from main sequence stars of types F to K.  Hence a mix of
all the template stars in Table \ref{tab-templates} would seem
appropriate.

A remaining potential problem is that the macroturbulent
broadening in our template stars may differ from the macroturbulence
characteristics of the cluster stars.  This difference may be of the
order of 1-3 \kms\ for giants and supergiants (Gray \cite{Gray81};
Gray and Toner \cite{GT87}) and somewhat larger for main sequence
stars (Gray \cite{Gray84}).  Since giants and supergiants are likely
to dominate the line widths in both clusters, the importance of
macroturbulence will be small, and will not be considered further.

\subsubsection{Selecting the optimum wavelength region}
\label{wave}

We used {\tt fxcor} to carry out separate cross-correlations for a
number of different wavelength regions.  

Prior to cross-correlation, we masked out
strong emission lines, telluric absorption features, and absorption in
the \nai\ D lines (see discussion in section \ref{nad} below).  The
spectrum of \#23 shortward of 4745 \AA\ is contaminated by nebular
emission and early-type stars. For cluster \#34, this part of the
spectrum has in addition poor signal-to-noise ratio.  We therefore
omitted this spectral region from the cross-correlation analysis.

   \begin{table}
   \begin{center}
      \caption[]{
Spectral regions used for cross-correlation in this paper in the rest
frame of ESO\,338--IG04.  `G' and `R' refer to the short and long
wavelength detectors, respectively, of the UVES red arm, avoiding the
strongest emission lines and telluric absorption.  Weaker emission
lines remaining have been masked out. The subscripts `B' and `R'
correspond to the `blue' and `red' parts of each region, respectively.
}
      \label{tab-regions}
      \begin{tabular}{ll}
\hline
Rest wavelength range (\AA) & Name \\
\hline

4606-4952, 5023-5552&  G  \\
5023-5552           &  G$_{\rm R}$ \\
5665-6273, 6320-6544&  R  \\
5665-6273           &  R$_{\rm B}$   \\
6320-6544           &  R$_{\rm R}$  \\
\hline
      \end{tabular}
   \end{center}
\end{table}

We show the names and ranges of the spectral regions we have used
in Table \ref{tab-regions}. Inspection of the spectra (see for example
Figure \ref{f-refstars}) showed that those features which are visible
above the noise in the cluster spectra tend also to be present in all
our template stars, though with different strengths.  All the
template stars gave a positive cross-correlation signal between
4606 \AA\ and 6545 \AA, and always at the same radial velocity, to
within $\pm 4$ \kms.  The blue part of region G (4606-4952 \AA)
gave less reliable results, and will not be further considered.
We found that the best cross-correlations were obtained if we filtered
out the lowest frequencies in the spectrum.  For this we used a ramp
filter rising from zero to unity at wavenumber 90 (40) for spectral regions
G and G$_{\rm R}$ (R$_{\rm B}$, R and R$_{\rm R}$).

\begin{figure}
 \resizebox{\hsize}{!}{\includegraphics[bb=145 170 300 400]{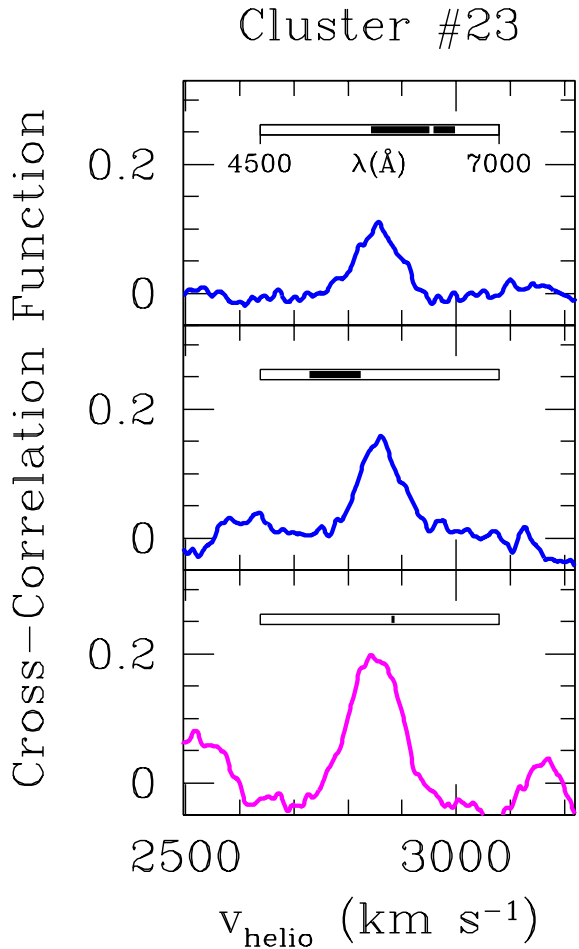}}

      \caption{
      Examples of cross-correlation functions of cluster \#23, here
      using the K4 supergiant HR~6961 as template star, in
      spectral regions R (top) and G$_{\rm R}$ (middle; see Table
      \ref{tab-regions} for details) and in the region around the
      \nai\ D lines towards \#23 (bottom).  The ranges in rest
      wavelength used for each cross-correlation are shown by the
      horizontal bar in each plot: the bar stretches from 4500 \AA\ to
      7000 \AA. }
\label{f-ccf23} 
\end{figure}

   \begin{figure}
 \resizebox{\hsize}{!}{\includegraphics[bb=145 170 300 350]{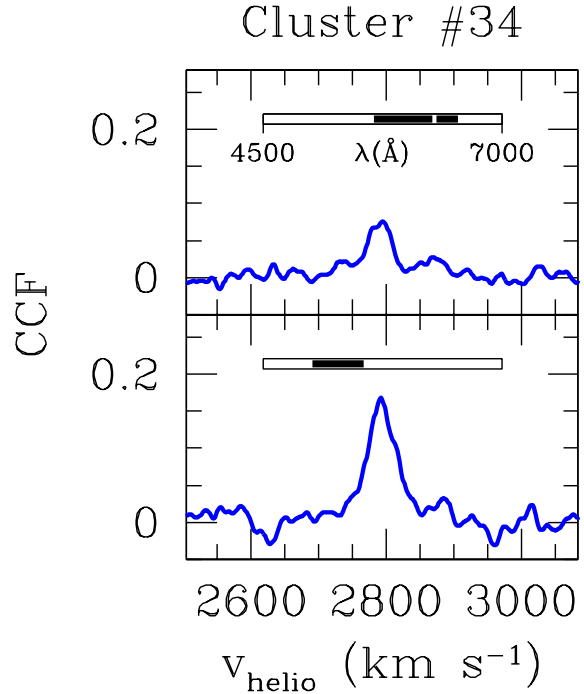}}
      \caption{
      Same as Figure \ref{f-ccf23}, but for
      cluster \#34 using the same template star. The upper panel
      shows spectral region R, the lower panel G$_{\rm R}$.}
\label{f-ccf34} 
\end{figure}

For both clusters, the highest cross-correlation signal was obtained 
for spectral region G$_{\rm R}$, irrespective of which template star was used.
Figures \ref{f-ccf23} and \ref{f-ccf34} show examples of
cross-correlation functions (CCFs) for one of the template stars,
HR~6961. These figures illustrate how the CCF peak amplitudes
differ between wavelength regions.  The CCF peak amplitudes for region
G$_{\rm R}$ are always significantly larger than regions G, R$_{\rm
B}$ and R$_{\rm R}$, which have peak values similar to that for R.

The arguments for the choice of template star presented in Sect.\
\ref{refstars}, based on the analysis using Starburst99, are of course
best founded for wavelength regions that include some of the Lick indices (Worthey \etal\
\cite{Worthey}). This is fulfilled for region G$_{\rm R}$, about half of whose
wavelength range is covered by these
indices. The other wavelength regions defined in Table
\ref{tab-regions} are clearly less suitable from this point of view.
While region R$_{\rm B}$ contains five defined indices, two are TiO
and one is Na~D, which we have masked out to avoid contamination by
interstellar absorption. The TiO lines could be problematic since they
are strong only in very cool stars with later spectral type than any
of our template stars, but which may nevertheless be present in our
clusters (\eg\ M giants and supergiants).
In summary, region G$_{\rm R}$ appears to be the safest to use.  It is
also the wavelength region which contains most of apparent absorption
features and for which the strength of the cross-correlation data is
clearly the strongest.

\subsubsection{Calibrations and uncertainties}
\label{ccf}

For each cluster, spectral region and template star, we measured
the heliocentric radial velocity ($v_{\rm helio}$), the amplitude
($h$) and full width at half maximum ($W_v$) of the best Gaussian fit
to the peak of the cross-correlation function.  To estimate $\Delta
W$, the uncertainty in $W_v$, we made several measurements with {\tt
fxcor} of the width of each CCF peak, varying within plausible limits
the baseline and velocity region for the fit, and taking the standard
deviation of several measurements.  Our notation is summarised in
Table \ref{notation}.

   \begin{table}
   \begin{center}
      \caption[]{Notation for cross-correlation analysis.}
      \label{notation}
      \begin{tabular}{ll}
\hline
Symbol & Definition \\
\hline

$v_{\rm helio}$ & Heliocentric radial velocity  \\
$W_v$        & Measured FWHM of the CCF peak  \\
$\Delta W $  & Estimated uncertainty of $W_v$   \\
$h$          & Peak value of the CCF   \\
$\Delta h$   & Tolerance in $h$ matching   \\
$\sigma $    & Velocity dispersion \\
$s_W$        & RMS scatter in $W_v$ of  derived $W_v-\sigma$ relation   \\
$s_\sigma$   & RMS scatter in  $\sigma$ of derived $W_v-\sigma$ relation \\
\hline
      \end{tabular}
   \end{center}    
\end{table}

We measured $v_{\rm helio}$ independently for each cluster and
template star, using accurate radial velocities from the literature
(Table \ref{tab-templates}).  Since the catalogue velocity of
HR~8299 was only accurate to 10 \kms, and no catalogue velocity was
available for HR 7706, we cross-correlated these with HR 6961 and
measured new heliocentric velocities for these stars (Table
\ref{tab-templates}).

To estimate the line-of-sight velocity dispersion $\sigma$ from width
of the CCF, we followed the method of Ho and Filippenko (\cite{HFa},
\cite{HFb}).    The basic procedure was as follows:
we first broadened each template star spectrum with Gaussians of known
$\sigma$, which we then cross-correlated with the corresponding
observed template star spectrum.  From the resulting CCF, we measured
the width ($W_v$) and amplitude ($h$) of the best Gaussian fit to the
peak, in the same way as for the cluster spectra, and used this to
calibrate a relation that gave the velocity dispersion $\sigma$ as a
function of the measured CCF width $W_v$.

Our clusters are however faint, and our CCFs peak at lower values than
those of previous authors.  We therefore adapted the method in order
to give a more careful calibration for our data.

\begin{figure}
 \resizebox{\hsize}{!}{\includegraphics[bb=30 150 550 690]{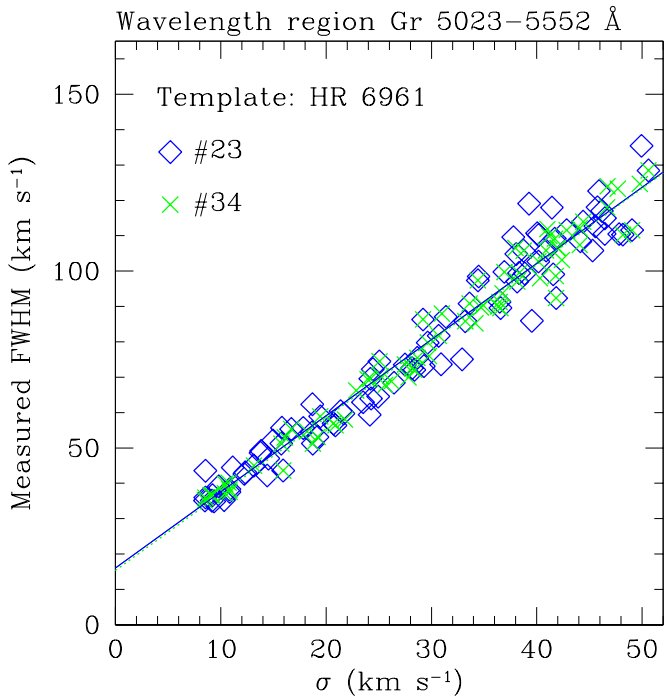}}
      \caption{
An example plot showing how we have estimated velocity dispersion
$\sigma$ by measuring the width of fits to the CCF peak.  Diamonds
represent models which match, to within $\pm0.015$, the observed CCF
peak amplitude for cluster \#23.  Crosses match the peak
amplitude for \#34. }

\label{f-calib} 
\end{figure}

To explore how the quality of the spectra affects the $W_v - \sigma$
calibration, we added different amounts of random, Gaussian noise to
the template star spectra using the {\sc iraf} task {\tt mknoise}.
Each of these noisy model spectra was then cross-correlated
with the corresponding observed template star spectrum.  From among
the resulting CCFs, we selected those whose peak amplitude $h$ 
matched that of the the observed cluster--template CCF, within a
tolerance $\Delta h$ (Figure \ref{f-calib}), and used those data for
our $W_v - \sigma$ calibration.

For CCF peak amplitudes comparable to those of our
observations, we found that the relation between fitted CCF width
$W_v$ and input $\sigma$ was always well-described by a linear
relation whose parameters vary with template star and wavelength
region.  The value of the heliocentric velocity was not affected
by low CCF peak amplitude, to within the errors. 

To obtain a realistic estimate of the errors in the velocity
dispersion calibration, we defined an acceptable spread $\Delta h$
about the measured CCF peak amplitude $(h_{\rm meas}^C\pm\Delta h)$,
and used only points lying in this interval. Here, $h_{\rm meas}^C$ is
the measured CCF peak amplitude for cluster $C$.  The value of $\Delta
h$ defines how closely we require the peak amplitude of our simulated
CCFs to match that of the CCF peak from cross-correlating the cluster
data with the template stars.  The uncertainty in the derived value of
$\sigma$ is clearly related to the scatter in each fit (see Figure
\ref{f-calib}), but it also depends on the value of $\Delta h$.  For
example, when $\Delta h$ is large, the scatter\footnote{We denote the
RMS scatter in $W_v$ and $\sigma$ about the linear relation as $s_W$
and $s_\sigma$, respectively.} about the linear relation between $W_v$
and $\sigma$ is also large.

For each combination of cluster, template star and spectral region, we
adjusted the value of $\Delta h$ until $s_W$, the scatter in the CCF width,
approached $\Delta W$, the uncertainty we estimated from using {\tt fxcor} 
to measure the same CCF width (\eg\ Figure \ref{f-calib}).  
In most cases, $s_W$ was somewhat larger than $\Delta W$, even for 
arbitrarily small values of $\Delta h$. Hence the uncertainty in the 
derived $\sigma$ values is not only due to uncertainty in measuring
$W_v$ but includes a small additional component from the 
cross-correlation method itself. % which may be of systematic nature.
Hence, $s_\sigma$ should give a realistic estimate of the 
uncertainty in each $\sigma$ measurement, and these are the values
quoted in parentheses in Table \ref{tab-sigmas}.

The derived velocity dispersions show some variation with template
star, though where the CCF amplitude is highest (region G$_{\rm R}$),
the derived values of $\sigma$ are gratifyingly similar. The F5~{\sc
v} template star HR 7749 gives systematically lower values of $\sigma$
for both clusters, but only towards the blue, where we expect more
contamination from early-type stars. For these reasons, and those
outlined in Sect.\ \ref{wave}, we regard the blue part of the G window
as less reliable and prefer therefore the results from region G$_{\rm
R}$.  The higher measured dispersion in R$_{\rm R}$ could be due in
part to contamination from the broader lines seen in for example
Si\,{\sc ii} in some early-type stars ({\sc uvespop} database, Bagnulo
\etal\ \cite{Bagnulo}). However, a velocity dispersion that varies as
a function of wavelength is not necessarily unphysical but may arise
naturally due to mass segregation, although we do not claim that this
is the reason for the present discrepancy.

As we argued in Sect.\ \ref{refstars}, early-type supergiants are
not likely contributors to the absorption spectra in the visual region
of either \#23 or \#34.  We nevertheless carried out some tests with
an A-type supergiant as template star, in order to further constrain
the dependence of our results on the assumed template star. The
A2\,{\sc I}{\small ab} star HD~102878 from the {\sc uvespop} database
(Bagnulo \etal\ \cite{Bagnulo}) was used for this purpose.  We found
that while this star gave acceptable cross-correlation signals, indicating the same
radial velocities as for our standard set of templates, we were unable
to use it for obtaining a robust $\sigma - W_v$ calibration
relation for our noisy data.  Better results were obtained, however,
for a hybrid template star which we constructed by adding HD~102878 and
the K4 supergiant HR~6961 in proportions 1:4. This hybrid star
produced radial velocities and $\sigma$ fully consistent with those of
our normal template stars. Hence, even if early-type stars could
contribute to the continua, they would not contribute much to the CCF
signal.

   \begin{table*}
      \begin{center}
      \caption[]{
Derived velocity dispersions in \kms\ for the two clusters, for each
template star and spectral region.  Errors are 1-sigma and are equal 
to the RMS scatter in the $\sigma - W_v$ relation for each
particular measurement, see text for details.  

}
      \label{tab-sigmas}

      \begin{tabular}{llccccc}

\hline
ref.\ & Type/ &  \multicolumn{5}{c}{Spectral region (see Table \ref{tab-regions})}\\ 
star & class                    & G          & G$_{\rm R}$         & R$_{\rm B}$         & R          & R$_{\rm R}$         \\% & Na \\
\hline
\multicolumn{7}{c}{Cluster \#23}\\
\hline
HR~6961  & K4\,{\sc i}{\small b} & 33.4 (2.5)  & 32.3 (2.2)  & 27.4 (5.1)  & 36.2 (3.1)  & 38.6 (2.8) \\
HR~7277  & K1\,{\sc i}           & 33.0 (3.2)  & 32.8 (2.4)  & 29.7 (6.7)  & 37.2 (2.9)  & 48.6 (7.8) \\
HR~7706  & K1\,{\sc iii}         & 32.5 (3.1)  & 31.3 (3.1)  & 29.3 (3.6)  & 33.2 (4.8)  & 41.1 (4.4) \\
HR~7749  & F5\,{\sc v}           & 21.2 (2.7)  & 32.0 (2.4)  & 30.9 (3.5)  & 39.7 (3.9)  & 46.0 (6.2) \\
HR~8299  & G5\,{\sc iii}         & 35.4 (2.1)  & 34.6 (3.3)  & 39.5 (6.1)  & 37.7 (2.4)  & 43.8 (4.6) \\
\hline
Hybrid$^1$ & K+A & 31.9 (3.4) & 29.9 (3.5) & 29.4 (5.0) & 30.2 (2.4) & 37.5 (13) \\
\hline \\
\multicolumn{7}{c}{Cluster \#34}\\      
\hline  
HR~6961  & K4\,{\sc i}{\small b} & 18.6 (2.4)  & 17.4 (2.0)  & 19.1 (2.8)  & 15.8 (3.0)  & 34.3 (19)  \\
HR~7277  & K1\,{\sc i}           & 18.2 (2.4)  & 17.6 (1.7)  & 16.2 (4.4)  & 11.0 (5.6)  & 18.2 (4.4) \\
HR~7706  & K1\,{\sc iii}         & 18.0 (2.5)  & 16.8 (3.0)  & 16.3 (3.5)  & 15.8 (3.5)  & 15.8 (8.0) \\
HR~7749  & F5\,{\sc v}           &  9.1 (1.9)  & 17.4 (2.0)  & 16.0 (3.5)  & 22.1 (3.3)  & 55.8 (30) \\
HR~8299  & G5\,{\sc iii}         & 18.8 (1.7)  & 18.1 (1.8)  & 17.6 (8.2)  & 18.1 (2.7)  & 20.3 (32) \\
\hline
Hybrid$^1$ & K+A  &18.5 (2.3)  &18.4 (2.4)  &15.6 (5.2)  &15.3 (2.4)  &18.5 (15) \\
\hline
      \end{tabular}
      \end{center}
\noindent Note: (1) K4\,{\sc i}{\small b} and A2\,{\sc I}{\small ab} in proportion  4:1.
    \end{table*}

   \begin{table*}
      \begin{center}
      \caption[]{
Derived heliocentric velocities in \kms\ for the two clusters, for
each template star and spectral region.  Errors are formal errors in
Gaussian fitting to the CCF peak from {\tt fxcor}.  

}
      \label{tab-velocities}
      \begin{tabular}{llccccc}
\hline
ref.\ & Type/ &  \multicolumn{5}{c}{Spectral region (see Table \ref{tab-regions})}\\
star & class                    &  G       & G$_{\rm R}$       & R$_{\rm B}$       & R        & R$_{\rm R}$       \\% & Na \\
\hline
\multicolumn{7}{c}{Cluster \#23}\\
\hline
HR~6961 & K4\,{\sc i}{\small b} & 2860 (7) & 2860 (4) & 2858 (7) & 2857 (7) & 2855 (7) \\% & 2846 (7) \\
HR~7277 & K1\,{\sc i}           & 2857 (7) & 2858 (4) & 2855 (6) & 2854 (7) & 2853 (7) \\% & 2843 (6) \\
HR~7706 & K1\,{\sc iii}         & 2860 (8) & 2861 (5) & 2857 (9) & 2856 (9) & 2856 (9) \\% & 2848 (6) \\
HR~7749 & F5\,{\sc v}           & 2858 (8) & 2860 (5) & 2857 (8) & 2854 (9) & 2853 (8) \\% & 2845 (6) \\
HR~8299 & G5\,{\sc iii}         & 2859 (7) & 2860 (4) & 2857 (7) & 2855 (7) & 2854 (7) \\% & --- \\ % 2842 (7) \\
\hline
Hybrid$^1$ & K+A & 2860 (7) & 2859 (4) & 2857 (8)  & 2857 (7)  & 2856 (7) \\
\hline \\
\multicolumn{7}{c}{Cluster \#34}\\      
\hline        	         
HR~6961 & K4\,{\sc i}{\small b} & 2794 (4) & 2795 (2) & 2793 (5) & 2792 (5) & 2789 (14)\\% & 2766 (22) \\
HR~7277 & K1\,{\sc i}           & 2792 (4) & 2793 (2) & 2791 (5) & 2790 (5) & 2789 (5) \\% & --- \\ %2790 (6) \\
HR~7706 & K1\,{\sc iii}         & 2794 (4) & 2796 (2) & 2793 (7) & 2793 (5) & 2793 (5) \\% & 2788 (7) \\
HR~7749 & F5\,{\sc v}           & 2793 (4) & 2794 (2) & 2793 (7) & 2790 (8) & 2785 (13)\\% & 2800 (3) \\
HR~8299 & G5\,{\sc iii}         & 2794 (3) & 2795 (2) & 2793 (6) & 2793 (6) & 2792 (5) \\% & --- \\ %2800 (4) \\
\hline        	         
Hybrid$^1$  & K+A &  2794 (4) & 2795 (2) & 2792 (5)&  2792 (5) & 2791 (5) \\
\hline        	         

      \end{tabular}
      \end{center}
\noindent Note: (1) K4\,{\sc i}{\small b} and A2\,{\sc I}{\small ab} in proportion  4:1.
    \end{table*}

\subsubsection{Results for cluster \#23}
\label{results}

In Tables \ref{tab-sigmas} and \ref{tab-velocities} we present
the resulting velocity dispersions and radial velocities for the
different wavelength regions and template stars.

For \#23, we find a weighted average value of $\sigma_{\rm 23} =32.5
\pm 2.7$ \kms\ in region G$_{\rm R}$ (Sect.\ \ref{wave}), when using
all 5 template stars, and $\sigma_{\rm 23} =32.5 \pm 2.3 $ if we
only use the two K supergiants (see discussion in Sect.\
\ref{refstars}).  The uncertainty here includes both the uncertainty in
each measurement ($s_\sigma$; the values quoted in parentheses in
Table \ref{tab-sigmas}, see Sect.\ \ref{ccf}), and the smaller
variation between results for different template stars (\ie\ the standard
deviation of the $\sigma$ column in question). We use this method to
estimate the total uncertainty in $\sigma$ in what follows.
For region R we find $\sigma_{\rm 23} =37.1\pm 3.4$ \kms\ when using
all stars and $\sigma_{\rm 23} =36.7\pm 3.1$ when using the K
supergiants only.  Combining the measurements from G$_{\rm R}$ and R
we obtain $\sigma_{\rm 23} =34.4\pm 3.8$ \kms\ when using all stars,
and $\sigma_{\rm 23} =34.1\pm 3.5$ \kms\ when using only the K
supergiants.

Following the discussion in Sect \ref{refstars} giving higher weight
to the K super giants, and noting the better quality of the G$_{\rm R}$ data,
we adopt as our best estimate $\sigma_{\rm 23} =32.5 \pm 2.5$ \kms.
We note that adopting a velocity dispersion of the magnitude implied
from region R would produce a virial mass that is $25\pm25$\% higher.

We find heliocentric velocities $v_{\rm helio,23}=2859.7$ \kms\ in
G$_{\rm R}$ and $2855.2$ in R, with a scatter between the five
template stars that in both cases is only $\sim 1$ \kms.
Restricting to K supergiants only makes no significant difference for
$v_{\rm helio,23}$.  The output uncertainties on $v_{\rm helio}$ from
{\tt fxcor} are on the order of 4 \kms\ in G$_{\rm R}$ and 8 \kms\ in R (see
Table \ref{tab-velocities}). The small star-to-star scatter suggests
that the uncertainty is dominated by the quality of our cluster
spectra rather than by template star mismatches.
We note a systematic difference of $\sim 4$ \kms\ between regions
G$_{\rm R}$ and R, that however is within the uncertainties. We base
our final value on the G$_{\rm R}$ data: $v_{\rm helio,23}=2859.7 \pm
4.4$.

To investigate possible contamination from other clusters, we compared
cross-correlation results for the 1.8-arcsec extraction with the
narrower, 1-arcsec wide, extraction around cluster \#23 (see Sect.\
\ref{sec-spatial}). The narrower extraction gives CCFs with poorer
signal-to-noise ratio but very similar velocity dispersion in both G$_{\rm R}$
and R.  Also, the use of hybrid templates made from HD~102878 (A
supergiant) and HR~6961 (K supergiant) produced very similar results
($\sigma_{\rm 23} =30$ \kms\ in both G$_{\rm R}$ and R).

\subsubsection{Results for cluster \#34}

For cluster \#34 (Tables \ref{tab-sigmas} and
\ref{tab-velocities}) the analysis was more straightforward.  We
measured $\sigma_{\rm 34}=17.6\pm2.0$ \kms\ in G$_{\rm R}$ and $17.4 \pm 4.4$,
in R. Combining G$_{\rm R}$ and R yields $\sigma_{\rm 34}=17.5\pm2.8$ \kms.

For the radial velocities we find $v_{\rm helio,34}=2794.6\pm2.2$ \kms\ 
in G$_{\rm R}$ and $2791.7\pm5.6$ in R, i.e. we see a similar systematic difference
between G$_{\rm R}$ and R as for cluster \#23, but which again is within the
uncertainties.   

Given the better quality of the G$_{\rm R}$ data, we adopt these results for 
$\sigma_{\rm 34}$ and $v_{\rm helio,34}$.

\subsection{Evidence for ISM gas flows from Na~{\sc i} lines}
\label{nad}

We also carried out cross-correlations on the narrow spectral region
5874-5912 \AA, which is dominated by the \nai~D doublet. For cluster 
\#23 this was successful (see Fig.\ \ref{f-ccf23}), but the wavelength region was too short to
yield anything useful in the more noisy spectrum of \#34.  Nor
could we obtain a reliable velocity dispersion calibration.  The
result would in any case be difficult to interpret, since we might also
expect a contribution from ISM absorption.
The heliocentric velocity is well-determined and remarkably low.  At
2845$\pm$7 \kms\ it is 25 \kms\ lower than that inferred from region
G$_{\rm R}$ along the same line of sight.  Presumably, the \nai\
absorption we are seeing here is a combination of both stellar lines
and cold interstellar medium moving at a speed of $\sim20$ \kms\ along
the line of sight to the cluster (cf.\ Schwartz and Martin
\cite{sm}).  The emission lines also show evidence for a blue-shifted
component (see Sect.\ \ref{emlines}).

\subsection{Cluster sizes}

We have measured the spatial extent of the two clusters using the HST
images described and presented by {\"O}stlin et al.\ (\cite{OBR}).

To measure the sizes of \#23 and \#34 we used curves of growth
extending from 0 to 5 Planetary Camera (PC) pixels.  One PC pixel has
a size of 0\arcsec.0455 which corresponds to 8.27 pc at the adopted
distance of ESO\,338--IG04.  We used observations in 4 passbands:
F336W, F439W, F555W and F814W.

The growth curves were compared to model PSFs generated with Tiny
Tim\footnote{http://www.stsci.edu/software/tinytim} (Krist and Hook
2003) that we convolved with Gaussian profiles and King profiles with
a concentration parameter $c=2$.

Using the whole growth curve from 0 to 5 pixels, the resulting sizes
are well-constrained.
We find the best-fitting effective half-light radii to be $r_e= 5.2\pm1.0$
and $5.6\pm0.4$ pc for \#23 and \#34, respectively, when using a
Gaussian profile.  Since the difference in estimated effective radius
is in general small when comparing Gaussians and different King model
assumptions, $r_e$ is a quite model-independent, and hence robust, measure
of the size of a cluster (Carlson and Holtzman \cite{CH}; Maraston
\etal\ \cite{Maraston}).

\subsection{Virial masses}
\label{masses}

From the virial theorem, we can estimate the mass of a cluster if we
know its velocity dispersion $\sigma$ and its size.  Assuming the
cluster is in virial equilibrium, its mass is given by

$$%\begin{equation}
M_{\rm vir} = k \, \sigma^2 \,r_m/G, 
$$%\end{equation}

\noindent (e.g.\ Spitzer \cite{spitz}, Smith and Gallagher \cite{SG}, 
McCrady \etal\ \cite{MGG}) where $r_m$ is the half-mass radius. Taking
$k=7.5 , r_m=\frac{4}{3}r_e$ (Spitzer \cite{spitz}), we find that the
cluster masses are given by:

$$%\begin{equation}
 M_{\rm vir} = 2.324 \times10^3 ~\left ( \frac{\sigma}{{\rm \kms}} \right )^2 \left(\frac{r_e}{{\rm pc}}\right) \msun
$$%\end{equation}

\noindent Specifically, for \#23,

$$%\begin{equation}
M_{23} = 1.28\times10^7 \,\left(\frac{\sigma_{23}}{32.5\ {\rm \kms}}\right )^2 \,\left(\frac{r_e}{5.2\ {\rm pc}}\right) \msun
$$%\end{equation}

\noindent and for \#34,

$$%\begin{equation}
M_{34} = 4.03\times10^6 \,\left(\frac{\sigma_{34}}{17.6\ {\rm \kms}}\right )^2 \,\left(\frac{r_e}{5.6\ {\rm pc}}\right) \msun
$$%\end{equation}

\noindent 
In Table \ref{properties} we summarise the properties of clusters \#23
and \#34, including the resulting dynamical mass-to-light ratios ($M/L_V$). 
The uncertainty intervals follows from a root-square addition of
the uncertainties in $\sigma$ and $r_e$.  Boily \etal\ (\cite{boily})
discuss the evolution of $\eta = k \cdot r_m/r_e$ with time during the
evolution of a cluster. For a cluster of the age of
\#23, \ie\ $\sim$ 5 Myr, Boily \etal\ (\cite{boily}) find $\eta \approx 10$,
which is what we have used above. For ages greater than 50 Myr,
however, they predict $\eta \approx 20$ for models with high initial
density.

   \begin{table*}
      \centering
      \caption[]{Properties of clusters
      23 and 34.  The colours, absolute magnitudes and mass--to--light ratios 
      are corrected
      for Galactic reddening according to Schlegel \etal\ (\cite{schlegel}), but
      not for internal reddening.  }
      \label{properties}

      \begin{tabular}{lcc}

Cluster no.\ &  \#23 & \#34 \\
\hline
Absolute magnitude $M_v$  & $-$15.5 & $-$12.8 \\
Colour $(v-i)$  & 0.11 & 0.86 \\
Effective radius $r_e$ (pc)      & $5.2\pm1.0$    & $5.6\pm0.4$ \\
Heliocentric velocity $v_{\rm helio}$ (\kms) & $2859.7\pm 4.4$  & $2794.6\pm2.2$ \\
Velocity dispersion $\sigma$ (\kms) & $32.5\pm2.5$ & $17.6\pm2$ \\
Virial mass ($10^6$\msun) & $13\pm3$ &  $4.0\pm 1$ \\
Mass-to-light ratio $(M/L_V)_\odot$ & $0.093\pm0.02$ & $0.35 \pm 0.09$ \\

\hline					   
      \end{tabular}
    \end{table*}

   \begin{figure*}
 \resizebox{\hsize}{!}{\includegraphics{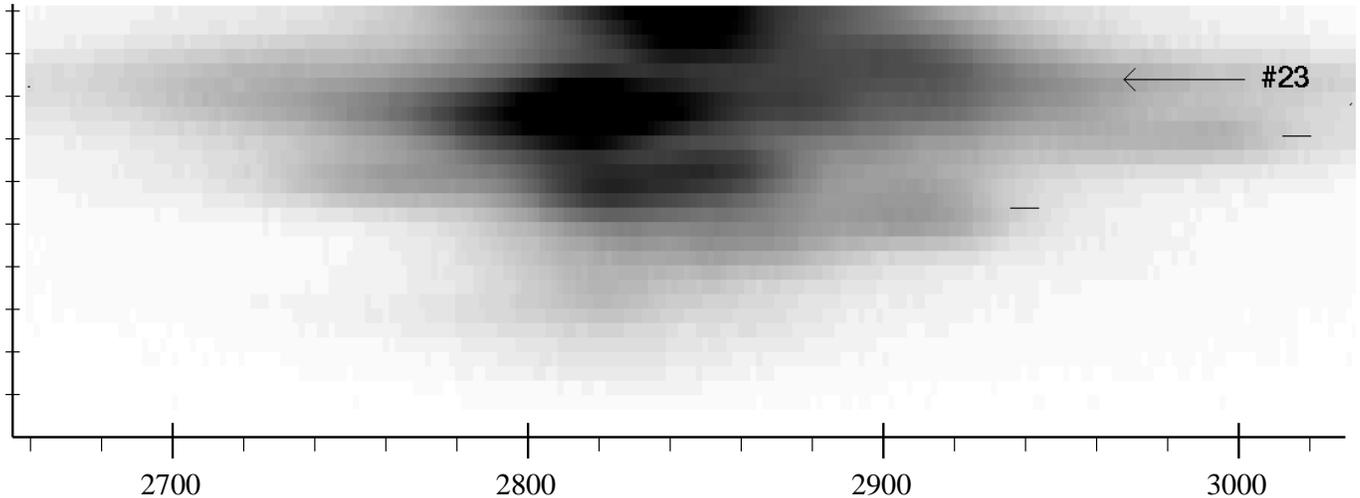}} 
 \caption{Two-dimensional spectrum around \oiii\ $\lambda$5007 showing 
 the complex 
 velocity structure in the ionised gas. The x-axis is labelled with
 velocity in units of \kms. The y-axis is the location along the slit,
 and the tick marks have a spacing of one arcsecond. The position of
 \#23 is indicated by an arrow, and two high velocity structures are
 marked by short dashes.}
\label{o3} 
\end{figure*}

\subsection{Velocity structure in emission lines}
\label{emlines}

The slit covers a region of strong emission from ionised gas. In Figure
\ref{o3} we show a spectral image centred on \oiii\ $\lambda$5007 
showing that the ionised gas motions along the slit are complex.  The
same features are seen in \ha.  The wavelength of the peak of the
\ha\ and \oiii\ $\lambda$5007 lines vary by almost 100 \kms\ along
the slit, but not in a monotonic fashion.  Closer inspection reveals
the presence of several velocity components at each position along the
slit.  For most positions at least three Gaussians appear to be needed to 
adequately reproduce the observed emission.

\begin{figure*}
 \resizebox{\hsize}{!}{\includegraphics[angle=-90]{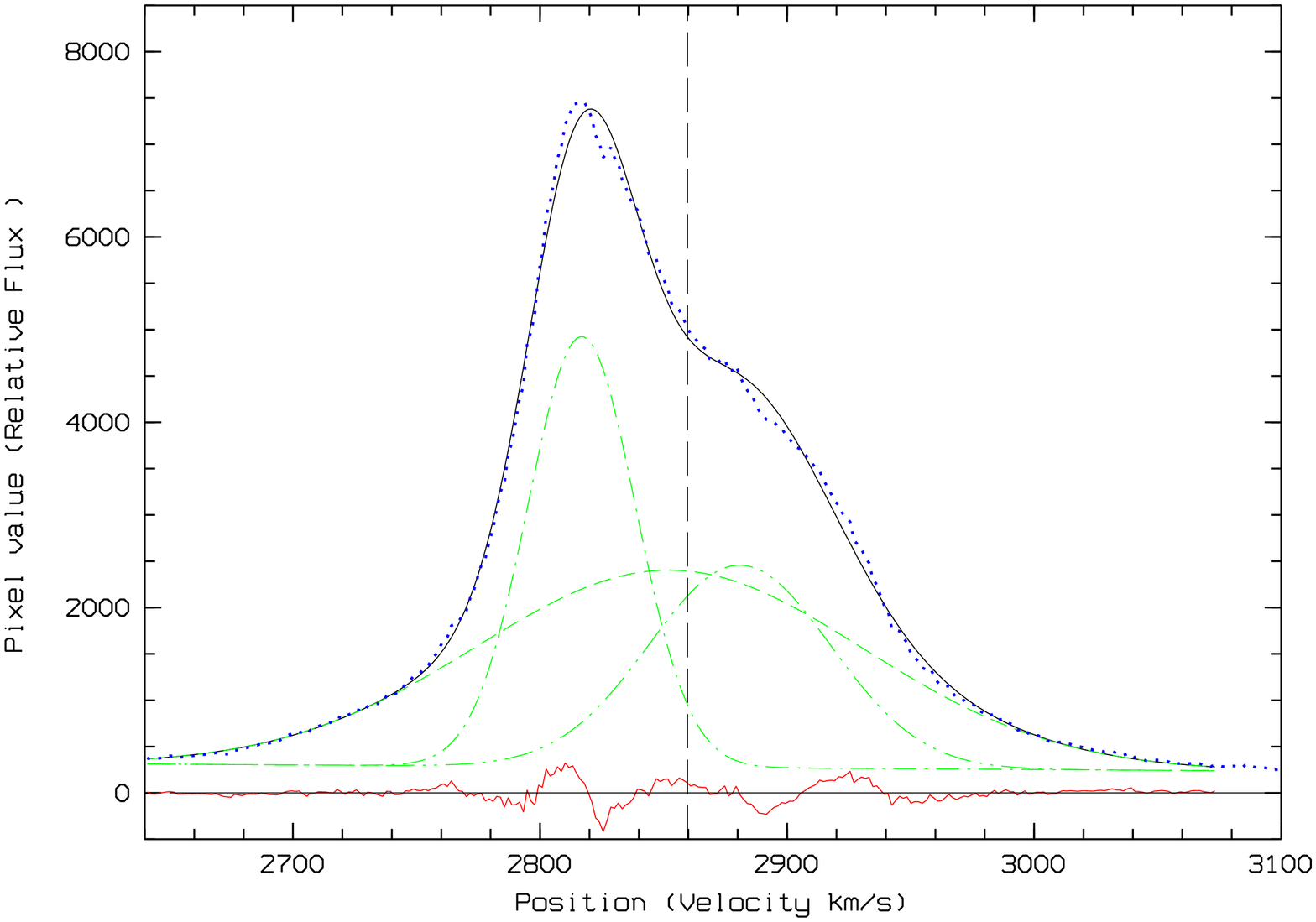}}

\caption{H$\alpha$ line profile 
for the two scan lines centred
on \#23, without background subtraction. The x-axis is labelled in
velocity. The thin (black) solid line shows the observed spectrum.
The (blue) dotted line shows the modelled spectrum which here is a sum
of 3 Gaussian profiles, which are shown as (green) dashed, dash-dot
and dash-dot-dot lines. The (red) dotted line at the bottom
shows the residuals (observed $-$ model spectrum). The vertical
(black) long-dashed line shows the velocity of \#23 determined from
absorption lines through cross-correlation.  This spectrum shows both
that there is both a blueshifted and a redshifted component, and that
there is a broad component with velocity close to the velocity of
\#23.}
\label{gauss} 
\end{figure*}

Figure \ref{gauss} shows the \ha\ line profile at the location of 
\#23. The line is very broad and clearly non-Gaussian, with peak 
intensity blue-shifted by 45 \kms \ with respect to the velocity
inferred from the absorption lines in \#23.  We fit three Gaussians to
the observed profile and find, one narrow blueshifted component (at
$-$40 \kms with respect to the absorption line velocity), one
redshifted component (at +20 \kms), and one broad ($\sigma = 65$ \kms)
component with nearly the same velocity as \#23.  We see the same
pattern in both \ha\ and \oiii\ 5007 \AA.

Looking at the scan lines to the north-east of cluster \#23 (above
\#23 in Fig.\ \ref{o3}), the line is split into two components: one
blueshifted and one redshifted with respect to the centre of the
cluster.  For the scan lines immediately below cluster \#23, 
the blueshifted component dominates.

The blue- and red-shifted components suggest a bubble expanding with a
velocity of $\sim$40 \kms.  If we take a characteristic radius of
1\arcsec\ (see Fig.\ \ref{o3} and Sect.\ \ref{23}), the timescale for
blowing such a bubble is $\sim$5 Myr, which is of the same order as
the age of \#23.  Our analysis here also shows that emission lines,
being much broader than the cluster's absorption lines, cannot
reliably be used to infer the dynamical mass of associated clusters
(cf.\ Turner \etal\ \cite{turner}).  A few scan lines further down
(marked with ticks in Fig.\ \ref{o3}) we see high-velocity components,
redshifted by 200 \kms\ and 80 \kms.  We finally note that cluster
\#34 has a stellar absorption line velocity which is 65 \kms\ less
than the ionised gas from the same position on the slit.

\section{Discussion: Comparing cluster masses and ages from photometry and spectroscopy}
\label{photmass}

The ages and photometric masses of the cluster population in
ESO\,338--IG04 have been investigated by {\"O}stlin \etal\
(\cite{OZBR}). Here, we discuss the properties of cluster \#23 and
\#34 in more detail, and compare with the virial mass estimates and
age estimates based on our UVES spectra.

For the sub-population of young (age $\le 20$ Myr) clusters, {\"O}stlin et
al.\ (\cite{OZBR}) preferred a low metallicity for both the stellar
and ionised gas components. This agrees with the nebular oxygen
abundance of $12+\log({\rm O/H})=8.0$ (Bergvall and {\"O}stlin \cite{BO}).
Moreover, to reproduce the youngest clusters, the IMF needs to have a
slope equal to or flatter than the Salpeter value($\alpha=2.35$) and
an upper mass limit $M_{\rm up} \ge 60 M_\odot$.  We shall refer to
this preferred metal-poor model as the ``standard''
model\footnote{Standard model: instant burst, IMF with slope
$\alpha=2.35$ (Salpeter) and mass range 0.08--120 \msun , gas covering
factor of unity, metallicity of stars and gas: $Z_{\rm gas}=0.002,
Z_{\rm stars}=0.001$.} in the following sections.

In the absence of dark matter, the photometric and dynamical masses
should agree. If this is not the case, the reason may be sought either
in the determination of the dynamical mass (because of mass
segregation, or non-equilibrium), or the photometric mass (IMF,
stellar parameters).  Below we discuss the photometric and dynamical
masses of each cluster in this context.

\subsection{Cluster \#23}   
\label{23}      

The standard model gives an age of 7 Myr, an internal reddening of
$E_{B-V}=0.05$ and a photometric mass of $5 \times 10^6 M_\odot$, \ie\
significantly lower than the virial mass. By adopting a steeper IMF
slope ($\alpha=2.85$) or a low upper mass limit ($M_{\rm up} = 20
M_\odot$), photometric masses in the range $2 {\rm ~to~} 6 \times 10^7
M_\odot$
can be produced.

The models of Zackrisson \etal\ (2001; Z01) that were used in {\"O}stlin
\etal\ (\cite{OZBR}) include nebular emission.  However, the nebular
emission produced by each cluster is assumed to be included in the
photometric aperture, which has a radius of 0\arcsec.14, or 25 pc at
the distance of ESO\,338--IG04.  This assumption appears to be valid
for the majority of young clusters, but it does not hold for \#23.
This cluster alone accounts for just over 10\% of the integrated
starburst flux at 2000 \AA\ and has ionised a large bubble (radius
$\sim 1$\arcsec, \ie\ nearly 200 pc), which, as we discussed in Sect.\
\ref{emlines}, is expanding (Figs. \ref{o3} and \ref{gauss}; see also
Fig.\ 2 in {\"O}stlin et al.\ \cite{OZBR}, or Fig.\ 7 in Hayes
\etal\ \cite{hayes}). Since ionised gas dominates the luminosity
output when the cluster is young (Bergvall and {\"O}stlin \cite{BO}), the
total luminosity, and hence the photometric mass of \#23 may be
underestimated by at most a factor of about two 
(see Fig.\ 17 in Bergvall and {\"O}stlin \cite{BO}).  For instantaneous
bursts of age $\sim 6$ Myr, the correction to the mass is instead of
the order $+25$\%. The displaced nebular emission also means that the
results of the photometric modelling for \#23 are more uncertain than
for most other young clusters in ESO\,338--IG04.

Indeed, recent \ha\ photometry with HST/ACS ({\"O}stlin \etal, in prep.),
gives an integrated \ha\ luminosity for \#23 (within the same aperture
as above) of $\sim 10^{32}$~W.  For the electron density derived from
our UVES spectra (Sect.\ \ref{spec23}), this implies a total ionised
gas mass of $\sim 10^{4}$ \msun\ or about one thousandth of the
dynamical mass. We caution that the density is derived from a larger
region where the nebular component is dominated by the more extended
gas around \#23, but the value is sufficiently small that we can
conclude that the mass of ionised gas in \#23 is negligible compared
to the dynamical and stellar masses. Moreover, the \ha\ emission line
equivalent width for \#23 derived from the ACS images is only about
$\sim 50$ \AA, much smaller than predicted for an age of $\sim 7$ Myr
(Starburst99, Leitherer \etal\ \cite{sb99}).  Hence, the ionising
photons produced in \#23 leak out over a much larger volume than the
size of the cluster.  Much of the ISM originally present in this
cluster is likely to have been removed already.

Another reason for the discrepant masses could be that the cluster is
not in virial equilibrium. This could be the case if the cluster is so
young that it has not yet had time to virialise, or because supernova
explosions have led to the prompt expulsion of gas from the cluster. In
such cases its dynamical mass may be overestimated by a factor of up to
3, depending on the star formation efficiency (SFE) in the cluster,
\ie\ the mass fraction of gas initially present that was turned into
stars (Bastian and Goodwin \cite{BGa}).  In the case of \#23 we would
infer a SFE of $\sim 40$\% for a stellar mass of $5\times10^6\msun$.
If we account for the displaced nebular emission, we obtain SFE
$>50$\%, the exact amount depending on the dynamical state of the
cluster prior to gas expulsion.  If the fraction of gas lost is larger
than 70\% the cluster may dissolve completely, but even a cluster
which remains bound will lose a few tenths of its initial stellar mass
(Bastian and Goodwin \cite{BGa}). Hence we expect \#23 to remain bound
but to lose a few tenths of its mass over the next $\sim$50 Myr.  

Alternative constraints on the age and IMF can be obtained from the
spectra presented in this paper.  Our analysis of Balmer and \hei\
absorption lines in Sect.\ \ref{balmer} indicated an age of
$6^{+4}_{-2}$ Myr. The effect of adopting an IMF with relatively few
massive stars is to increase the equivalent widths at young ages. A
low upper mass limit ($M_{\rm up} \ge 30 M_\odot$) may still be
marginally consistent with the data, but an IMF slope significantly
steeper than the Salpeter value predicts equivalent widths that are
too high (see for example Fig.\ 11 in Gonz{\'a}lez Delgado \etal\
\cite{GDLH}).

More evidence for the the presence of very massive stars comes from
the detection of intermediate-width stellar \heii\ $\lambda$4686 line
emission with equivalent width 0.8 \AA.  We see no obvious WR features
but interpret this line as a signature of Of stars, which are very
massive stars that may be on their way to becoming WR stars (Nota
\etal\ \cite{N96}; Schaerer and Vacca \cite{SV}). This observation
makes an IMF with a low upper mass limit improbable, but is consistent
with a Salpeter IMF and an age $\le 5$ Myr (Schaerer and Vacca
\cite{SV}; Leitherer \etal\ \cite{sb99}).

We noted in Sect.\ \ref{refstars} that the presence of strong calcium
triplet absorption from \#23 signals the presence of red
supergiants. For a Salpeter IMF and low metallicity, the presence of
RSGs indicates an age of $\ge 5$ Myr, whereas for an IMF with a low
upper mass limit, the RSGs require that the cluster is older than $10$
Myr. The simultaneous presence of RSGs and emission in \heii\
$\lambda$4686 is somewhat unexpected, but could be accounted for if
the formation of the cluster extended over a few Myr, rather than
being instantaneous.  Similar scenarios have been presented for
R136 (Massey and Hunter \cite{mh}) and NGC\,3603 (Pandey \etal\ \cite{pandey}).
Moreover, recent stellar models which include rotation predict
longer WR lifetimes (Meynet and Maeder \cite{md}).

Fellhauer and Kroupa (\cite{fk}) have proposed that the anomalously
massive, intermediate-age cluster W3 in NGC\,7252 may be the product
of merging of several less massive clusters.  Though the timescale for
this to happen seems short in the case of \#23, an earlier merger of
young clusters could lead to the presence of stellar populations with
different ages, and explain the simultaneous presence of Of stars and
RSGs.

It is also quite possible that stellar evolution in a cluster as massive as
\#23 is different compared to a lower-density environment.  Numerical 
simulations of massive star clusters indicate that dynamical friction
rapidly leads to strong mass segregation and possibly merging of the
most massive stars (\eg\ Portegies Zwart \etal\ 1999).  Massive-star
mergers would likely have a sizeable impact on the early spectral
evolution of such a star cluster.

In conclusion, we find strong support for an IMF of \#23 that extends to
high masses (on the order of 100 \msun) and has a slope not significantly 
steeper than the Salpeter value.
The cluster's photometric and dynamical masses are in reasonable
agreement, when allowing for the fact that most of its associated
nebular emission lies outside the photometric aperture, and the
possibility that the cluster is not yet fully virialised. A more
exotic possibility would be that the cluster contains of the order of
50\% dark matter.

\subsection{Cluster \#34}              %%%%%%%%%%%%%%%%%%%%%%%%%%%%%%%%%%%%%
\label{34}      

Cluster \#34 is much older than both \#23 and the current
starburst. Our standard$^8$ metal-poor model (considered in {\"O}stlin
\etal\ \cite{OZBR}, based on the code by Zackrisson \etal\ 2001;
Z01), implies an age of $\sim 1.5$ Gyr, and photometric mass
$2\times10^7$ \msun, almost an order of magnitude larger than the
virial estimate $\sim 4\times10^6$ \msun. Compared to the photometry
of \#34, this model is discordant at the 1-sigma level.  Varying the
IMF slope $\alpha$ in the range 1.35 to 2.85 gives comparable
qualities of fit and a mass $>10^7$ \msun. A better fit can
be obtained by adopting a model with significantly higher metallicity
($Z = 0.008$ to 0.04). This results in a lower age, 250 Myr (for
$Z=0.04$) to 500 Myr ($Z = 0.008$), and a somewhat lower mass of $
\sim 10^7 M_\odot$.
For all assumed metallicities and IMF parameters, we find a rather high
internal reddening, $E_{B-V}\approx 0.25$, which can be
compared to the low values for the galaxy's starburst centre
($E_{B-V}\le 0.05$; {\"O}stlin \etal\ 2003) suggesting that \#34 may lie
on the far side of ESO\,338-IG04.  About one third of the other old
and intermediate age clusters have similar implied reddenings.

In addition to Z01, whose stellar evolutionary tracks come mainly from
the Geneva group, we have used the spectral synthesis code {\sc
p{\'e}gase.2} (Fioc and Rocca-Volmerange 1997, 1999) which is mainly based
on tracks from the Padova group.  One effect of adopting a different
set of tracks is that {\sc p{\'e}gase.2} in general implies a lower
reddening and higher age for \#34. The best fit is again obtained for
a rather metal-rich ($Z=0.008$) model that gives an age of 1.2 Gyr, a
reddening of $E_{B-V}=0.1$ and mass $1.1\times10^7 M_\odot$ for a
Salpeter IMF.  However, {\sc p{\'e}gase.2} gives results that are
consistent with the photometry for all metallicities in the range
$0.002 < Z < 0.05$, with resulting ages from 0.8 to 1.2 Gyr, and
masses which differ by less than 20\%.  In summary, models based
on a Salpeter IMF predict a mass of $\sim 10^7 M_\odot$, irrespective
of the set of stellar evolutionary tracks or metallicity used.

All the mass estimates so far in this subsection assume a single power
law IMF, including remnants but excluding the gas returned in the
process of stellar evolution, which is assumed to have left the
cluster. If we instead assume an IMF like that seen in the solar
neighbourhood, with a flatter slope for stars with mass smaller than
$0.5-1$ \msun, the photometric masses quoted above are reduced by a
factor of $\sim 2$. For instance, a Scalo98 (Scalo \cite{Scalo98}) IMF
yields $\sim 5 \times 10^6 M_\odot$ for the best fitting models. 
This is only slightly larger than the virial mass and such an IMF is
also consistent with Galactic GCs (Chabrier and M{\'e}ra \cite{CM}).

The work by Boily \etal\ \cite{boily} suggests that for ages
larger than 50 Myr, masses may be underestimated by a factor of two if
the temporal evolution of $\eta$ is not accounted for (see also Sect.\
\ref{masses}). However, this is only true for clusters with high
initial mass surface density, and moreover, the evolution past 50 Myr
is not known.  Our \#34 has a present mass surface density of $\sim
2\times 10^4$ \msun/pc$^2$ (without any $\eta$-evolution included), so
strong effects are not expected.

Given the concordance of the photometric mass for different models and
metallicities, we conclude that the dynamical mass of \#34 is best
explained with an IMF where the low-mass part is similar to that
observed in the solar neighbourhood and in Galactic GCs.

\subsection{Implications for the globular cluster system in ESO\,338--IG04}

The two clusters studied in this paper are consistent with a normal
IMF, extending to low stellar masses ($M \sim 0.1 M_\odot$) and with a
slope in the high mass regime ($M > 1 M_\odot$) close to the Salpeter
value. Such IMFs will favour survival against cluster destruction
mechanisms related to stellar death and gas-expulsion.  A few SSCs in
other galaxies on the other hand seem to have rather odd IMFs,
\eg\ M82-F, for which a severe deficiency of low mass stars is implied
(Smith and Gallagher \cite{SG}), making its long-term survival unlikely.

All SSCs and GCs clusters line up nicely in a fundamental plane
(Walcher \etal\ \cite{walcher}) suggesting that initial conditions
regulate their properties (McLaughlin \cite{mclaughlin}).  That young
SSCs, including \#23 and \#34, follow the same trends suggests that
the same basic formation mechanism applies both for globular clusters in
ancient halos and for present-day mergers.  {\"O}stlin \etal\ (OZBR) used
the cluster age distribution in ESO\,338--IG04 to map its past star
and cluster formation history, and found that star clusters seem to make
up as much as several percent of the total stellar mass, in contrast
to the value of 0.3\% found for most galaxies (McLaughlin
\cite{mclaughlin}).  

Our study demonstrates the massive nature of at
least some of the star clusters in ESO\,338--IG04.  
More velocity
dispersion measurements in ESO\,338--IG04 would be illuminating.
Though they are the  brightest of the
subpopulations of young and intermediate-age clusters, respectively, 
the objects studied in this paper are just two of the hundreds of
clusters detected in this galaxy.  
Despite their impressive masses, \#23 and \#34 are apparently faint, with $m_v=17.4$ and 20.1, respectively.  Cluster \#34 is
the faintest
so far for which velocity dispersions have been derived from optical
spectroscopy. With the exception of two very massive intermediate age
clusters in NGC~7252 (Bastian \etal\ \cite{Bastian}), our clusters are
also the most distant yet probed.  In this paper we have shown that
velocity dispersion measurements of such faint clusters are indeed
feasible, and we expect that clusters one magnitude fainter would be
within reach with modest increase of integration time on VLT/UVES.
Several more clusters in ESO\,338--IG04 could be weighed by the same
method.

\section{Conclusions}

We have presented high resolution spectra of two luminous star
clusters in the luminous blue compact galaxy ESO\,338--IG04
(Tololo\,1924--416).  The spectra have been cross-correlated with
template stars observed using the same setup to determine each
cluster's line-of-sight velocity dispersion.  Using size estimates
from Hubble Space Telescope images, we have used the velocity
dispersions to determine the virial masses of the clusters.  Our mass
estimates have been compared to masses derived from HST photometry
fitted to spectral evolutionary synthesis models. This comparison
indicates that both clusters have rather normal IMFs, favouring their
survival against internal disruption mechanisms.

One of the clusters (\#23) is young, with an age of $\sim6-7$ Myr and
mass in excess of $10^7$ \msun, making it one of the most
massive very young clusters known. We find evidence for the
simultaneous presence of massive O-stars and red supergiants. The
\oiii\ $\lambda$5007 and \ha\ emission line profiles from the
region surrounding cluster \#23 indicate the presence of a bubble
expanding at a velocity $\sim40$ \kms.  In addition, we see signs of
neutral gas flows along the line of sight from the cross-correlation
analysis of the Na~{\sc i}\,D absorption lines.  The IMF of this
cluster shows no evidence for any deficiency of low-mass stars, but is
consistent with a Salpeter slope over the whole mass range 0.08--120
\msun. The inferred dynamical mass is larger than expected based on
the luminosity of the cluster, which may indicate that the cluster is
not yet virialised.

The other cluster (\#34) is older and has a mass of $4\times 10^6$
\msun.  A combined analysis with the photometric data suggests that it
has an age of between 0.3 and 1.4 Gyr.  The slope of the IMF at masses
less than 1 \msun\ must be flatter than the Salpeter value
for the virial and photometric masses to be consistent.  Such an IMF
is in agreement with that observed for old Galactic globular
clusters. This work confirms that this cluster is a bona-fide
young globular cluster.

\begin{acknowledgements}

We thank Andrea Modigliani for useful suggestions on the reduction of
UVES data.  The authors acknowledge financial support from the Swedish
research council and the Swedish National Space Board.  We thank
Matthew Hayes, Erik Zackrisson, Kambiz Fathi and Jan-Olov Persson for
useful discussions.

\end{acknowledgements}


\begin{thebibliography}{}

\bibitem[1995]{arp} Arp, H., Sandage, A., 1985, AJ, 90, 1163
\bibitem[2003]{Bagnulo} Bagnulo, S., Jehin, E., Ledoux, C., Cabanac, R., Melo, C., Gilmozzi, R., 2003, Messenger 114, 10
 \bibitem[2006]{Bastian} Bastian, N., Saglia, R. P., Goudfrooij, P., Kissler-Patig, M., Maraston, C., Schweizer, F., Zoccali, M., 2006, A\&A, 448, 881, astro-ph/0511033
 \bibitem[2006]{BGa} Bastian, N., Goodwin, S., 2006, MNRAS, 369, 9, astro-ph/0602465
 \bibitem[2002]{BO} Bergvall N., {\"O}stlin, G., 2002, A\&A 390, 891
 \bibitem[1994]{Bertelli} Bertelli, G., Bressan, A., Chiosi, C., Fagotto, F., Nasi, E. 1994, A\&AS, 106, 275
 \bibitem[1986]{bica} Bica, E., Alloin, D., 1986, A\&A 162, 21  
 \bibitem[2003]{BC} Bruzual, G., Charlot, S., 2003, MNRAS 344, 1000 
 \bibitem[2005]{boily} Boily, C.M., Lan{c}on, A., Dieters, S., Heggie, D.C., 
2005, ApJ, 620, L27 
 \bibitem[2001]{CH} Carlson, M. N., Holtzman, J. A., 2001, PASP, 113, 1522
 \bibitem[1997]{CM} Chabrier, G., M{\'e}ra, D., 1997, A\&A 328, 83
 \bibitem[1989]{Diaz} D{\'\i}az, A. I., Terlevich, E., Terlevich, R. 1989, MNRAS, 239, 325
 \bibitem[2000]{DL} Drilling, J. S., Landolt, A. U. 2000, in Cox A. N., ed., Allen's Astrophysical Quantities, 4th edn., Am. Inst. Phys., New York
\bibitem[2001]{drinkwater} Drinkwater, M.J., Gregg, M.D., Hilker, M., 2003, Nature, 423, 519
\bibitem[2000]{dolphin} Dolphin, A.E., 2000, PASP 112, 1397
\bibitem[2005]{fk} Fellhauer, M., Kroupa, P., 2005, MNRAS, 359, 223
\bibitem[1997]{fr97} Fioc, M., Rocca-Volmerange, B., 1997, A\&A 326, 950
\bibitem[1999]{fr99} Fioc, M., Rocca-Volmerange, B., 1999, astro-ph/9912179 ({\sc p{\'e}gase.2}) 
\bibitem[2002]{GG} Gilbert, A. M., Graham, J. R. 2002, in IAU
Symp.\ 207, Extragalactic Star Clusters, ed.\
E. K. Grebel, D. Geisler \& D. Minniti
  \bibitem[1997]{GD} Gonz{\'a}lez Delgado, R. M., Leitherer, C., Heckman,
T., Cervi{\~n}o, M. 1997, ApJ, 483, 705
  \bibitem[1999]{GDL99} Gonz{\'a}lez Delgado, R. M., Leitherer, C. 1999,
ApJS, 125, 479 (GDL99)
  \bibitem[1999]{GDLH} Gonz{\'a}lez Delgado, R. M., Leitherer, C., Heckman, T., 1999,
ApJS, 125, 489
  \bibitem[1981]{Gray81} Gray, D. F. 1981, ApJ, 251, 152
  \bibitem[1984]{Gray84} Gray, D. F. 1984, ApJ, 281, 719
  \bibitem[1987]{GT87}  Gray, D. F., Toner, C. G. 1987, ApJ, 322, 360
  \bibitem[2005]{hayes} Hayes, M., {\"O}stlin, G., Mas-Hesse, J.M., Kunth, D.,  
Leitherer, C., Petrosian, A., 2005, A\&A 438, 71, astro-ph/0503320
  \bibitem[1996a]{HFa} Ho, L. C., Filippenko, A. V., 1996a, ApJ, 466, 
L83
  \bibitem[1996b]{HFb} Ho, L. C., Filippenko, A. V., 1996b, ApJ, 472, 
600
  \bibitem[2004]{Kharchenko} Kharchenko, N.V., Piskunov, A.E., Scholz, R.-D., 2004, Astron. Nachr., 325, 439
  \bibitem[2003]{tinytim} Krist, J., Hook, R., 2003 ``The Tiny Time User's Guide'', version 6.1,
http://www.stsci.edu/software/tinytim
  \bibitem[2001]{Larsen01} Larsen, S. S., Brodie, J. P., Elmegreen
B. G., Efremov, Y. N., Hodge P. W., Richtler, T. 2001, ApJ, 556, 801
  \bibitem[2004]{Larsen04} Larsen, S. S., Brodie, J. P., Hunter, D. A. 2004, ApJ, 128, 2295
  \bibitem[1999]{sb99} Leitherer, C., Schaerer, D., Goldader, J.D., et al., 1999
ApJS, 123, 3 
  \bibitem[2001]{Maoz} Maoz, D., Ho, L. C., Sternberg, A. 2001, ApJ,
554, L139
  \bibitem[2004]{Maraston} Maraston, C., Bastian, N, Saglia R.P., Kissler-Patig M., Schweizer, F., Goudfrooij, P., 2004, A\&A 416, 467
\bibitem[1998]{mh} Massey, P., Hunter, D.A., 1998, ApJ, 493, 180
  \bibitem[2003]{MGG} McCrady, N., Gilbert, A. M., Graham, J. R., 2003,
ApJ, 596, 240
  \bibitem[2000]{mclaughlin} McLaughlin, D.E., 2000, ApJ 539, 618
  \bibitem[2002]{M} Mengel, S., Lehnert, M. D., Thatte, N., Genzel,
R., 2002, A\&A, 383, 137
  \bibitem[1995]{meurer} Meurer,  G.R., Heckman T.M., Leitherer, C., Kinney, A.,
Robert, C., Garnett, D.R., 1995, AJ, 110, 2665
  \bibitem[2005]{md} Meynet, G., Maeder, A., 2005, A\&A, 429, 581
  \bibitem[2002]{Nidever} Nidever, D. L., Marcy, G. W., Butler, R. P., Fischer, D. A., Vogt, S. S., 2002, ApJS, 141, 503
  \bibitem[1980]{Ochsenbein} Ochsenbein, F., 1980, Bull. Inf. CDS, 19, 74
  \bibitem[2000]{VizieR}  Ochsenbein F., Bauer P., Marcout, J., 2000, A\&AS, 143, 221  
  \bibitem[1996]{N96} Nota, A., Pasquali, A., Drissen, L., Leitherer,
C., Robert, C., Moffat, A. F. J., Schmutz, W. 1996, ApJS, 102, 383
  \bibitem[2001]{O01} Origlia, L., Leitherer, C., Aloisi, A., Greggio,
L., Tosi, M. 2001, AJ, 122, 815
  \bibitem[1998]{OBR} {\"O}stlin, G., Bergvall, N., R{\"o}nnback, J., 1998,  A\&A, 335, 85
  \bibitem[2001]{ostlin01} {\"O}stlin G., Amram P., Bergvall N., Masegosa J., 
Boulesteix J., Marquez I., 2001,  A\&A, 374, 800
  \bibitem[2003]{OZBR} {\"O}stlin, G., Zackrisson, E., Bergvall, N., R{\"o}nnback, J., 2003, A\&A, 408, 887, astro-ph/0306522
  \bibitem[2000]{pandey} Pandey, A.K., Ogura, K., Sekiguchi, K., 2000, PASJ 52, 847
  \bibitem[1999]{portegies} Portegies Zwart S.F., Makino, J., McMillan, S.L.W., Hut, P.,
1999, A\&A 348, 117
\bibitem [1955]{salpeter} Salpeter, E., 1955, ApJ, 121, 161
  \bibitem[1998]{Scalo98} Scalo, J. 1998, in ASP Conf.\ Ser., 142, The Stellar Initial Mass Function, 201
  \bibitem[1998]{SV} Schaerer, D., Vacca, W. D., 1998, ApJ, 497, 618
  \bibitem[1998]{schlegel} Schlegel, D.J., Finkbeiner, D.P., Davis M., 1998, ApJ, 500, 525
  \bibitem[2004]{sm} Schwartz, C. M., Martin, C. L., 2004, ApJ, 610, 201
  \bibitem[1974]{Simkin} Simkin, S.M., 1974, A\&A, 31, 129
  \bibitem[2001]{SG} Smith, L. J., \& Gallagher, J. S., III. 2001,
MNRAS, 326, 1027
  \bibitem[1987]{spitz} Spitzer, L., Jr., 1987, Dynamical Evolution of
Globular Clusters, Princeton Univ.\ Press, Princeton, NJ 
  \bibitem[1979]{TD} Tonry, J., \& Davis, M., 1979, AJ, 84, 1511
  \bibitem[2003]{turner} Turner, J. L., Beck, S. C., Crosthwaite, L. P., Larkin, J. E., McLean, I. S., Meier, D. S., 2003, Nature 423, 621
  \bibitem[1979]{U95} Underhill, A. 1995, ApJS, 100, 461
  \bibitem[1992]{VC} Vacca, W. C., \& Conti, P. S., 1992, ApJ, 401, 543
  \bibitem[2005]{walcher} Walcher, C.J., van der Marel, R.P., McLaughlin, D., et al., 2005, ApJ 618, 237
  \bibitem[2004]{Werk} Werk, J. K., Jangren, A., Salzer, J. I., 2004, ApJ, 617, 1004
  \bibitem[2003]{whitmore_rev} Whitmore, B.C., 2003, in ``A Decade of Hubble Space Telescope Science'', Eds. Livio M., Noll K., Stiavelli M., Space Telescope Science Institute symposium series, Vol. 14, p. 153,
Cambridge University Press, astro-ph/0012546
  \bibitem[1999a]{Whitmore} Whitmore, B., Heyer, I., Casertano, S., 1999, PASP, 111, 1559
  \bibitem[1999b]{W99b} Whitmore, B. C., Zhang, Q., Leitherer, C., Fall, M. S.,Schweizer, F., Miller, B. W., 1999, AJ, 118, 1551
  \bibitem[1994]{Worthey} Worthey, G., Faber, S. M., Gonzalez, J. J., Burstein, D. 1994, ApJ, 94, 687
  \bibitem[2001]{z01} Zackrisson E., Bergvall N., Olofsson K., Siebert A., 2001, 
A\&A 375, 814 (Z01)
\end{thebibliography}
\end{document}